\documentstyle[12pt]{article}

\renewcommand{\theequation}{\arabic{subsection}.\arabic{equation}}

\title{Positronium States in QED3} 
\author{T.\ W.\ Allen\ and C.\ J.\ Burden\\
{\small Department of Theoretical Physics,}\\
{\small Research School of Physical Sciences and Engineering,}\\
{\small Australian National University, Canberra, ACT 2601, Australia}\\}
\begin{document}
\maketitle
\abstract{The $e^-$-$e^+$ bound state spectrum of QED3 is investigated 
in the quenched ladder approximation to the 
homogeneous Bethe-Salpeter equation with fermion propagators 
from a Rainbow approximation Schwinger-Dyson solution with 
variable fermion mass.  
A detailed analysis of the analytic structure of the fermion 
propagator is performed so as to test the appropriateness of 
the methods employed.  The large fermion mass limit of the Bethe-Salpeter 
equation is also considered, including a derivation of the Schr\"{o}dinger 
equation,  and comparisons made with existing non-relativistic 
calculations.  \\
\\
PACS NUMBERS: 11.10.S, 11.10.K
\hfill\break}
\pagebreak
\subsection{Introduction}
The similarities between Quantum Electrodynamics in three space-time
dimensions (QED3) and Quantum Chromodynamics in four space-time
dimensions (QCD4) and the simplicity of the 
theory make QED3 attractive for the study of non-perturbative methods.  
QED3 is an abelian theory and provides a logarithmic
confining $e^-$-$e^+$ potential~\cite{BPR92}.

Our approach to positronium states in QED3 is via a solution to the 
homogenous Bethe-Salpeter equation with fermion propagator
input from the Schwinger-Dyson equation.
The full Schwinger-Dyson and Bethe-Salpeter equations are 
intractable. Here we consider a solvable system of integral 
equations within the quenched, ladder approximation.  This crude 
truncation of the full equations does break gauge covariance but 
has very attractive features and has been employed extensively
in QCD4 spectrum calculations~\cite{BSpapers,SC94}.

This study continues on from a previous study~\cite{Bu92} which uses
a four-component fermion version of QED3.  In this version, the massless case 
exhibits a chiral-like $U(2)$ symmetry broken into a 
$U(1) \times U(1)$
symmetry by the generation of a dynamical fermion mass, resulting in
a doublet of Goldstone bosons.  This pion-like solution is important
for drawing similarities between QED3 and QCD4.  The 
four component 
version of QED3 is also preferred to the two component 
version because 
the Dirac action in the two component version is not parity 
invariant for massive fermions.  QCD4 is parity invariant 
and we aim to have as much in common with
that theory as possible.

The previous work was restricted to zero bare fermion
mass, while in this study the bare mass is increased from
zero to large values in order to compare with results in the 
non-relativistic limit.  This study also takes a closer
look at the choice of fermion propagator input.  Knowledge
of the analytic properties of the fermion propagator is
important for determining the approximation's ability
to provide confinement and whether or not any singularities
will interfere with a Bethe-Salpeter
solution.    Based on the work of Maris~\cite{Ma93,Ma95} the 
occurrence of mass-like complex singularities
is expected which have the potential to influence our calculations.

In section 2 we look at the Bethe-Salpeter and Schwinger-Dyson 
approximations used in this work and the method used to find the
bound state masses.  A brief review of transformation properties in 
QED3 is given in the appendix. 
These transformation properties are of vital importance for 
an understanding of the structure of the $e^-$-$e^+$ vertex 
function and the classification of the bound states.
Section 3 describes the non-relativistic limit and the connection 
between the Bethe-Salpeter and Schr\"{o}dinger equations for QED3.  

In section 4 the approximation to the fermion propagator is 
detailed. The structure of the propagators will be analysed in 
the complex plane where we attempt to locate the expected 
mass-like singularities.  
In section 5 the Bethe-Salpeter
solutions are reported and comparisons are made with
non-relativistic limit calculations.  The results are discussed 
and conclusions given in section 6.

\setcounter{equation}{0}
\subsection{Solving the Bethe-Salpeter Equation}
The Bethe-Salpeter (BS) kernel for this work is a simple one-photon
exchange (ladder approximation) which is a commonly used starting 
point.  For convenience we use the quenched approximation,
work in Feynman gauge and work only with the Euclidean metric.
Fig.~1 shows the Bethe-Salpeter equation in the quenched ladder 
approximation.  The corresponding integral equation is
\begin{equation}
\Gamma(p,P) = -e^2 \int \mbox{$ \, \frac{d^3q}{(2\pi)^3} \,$} 
D(p-q) \gamma_\mu S(\mbox{$ \, \frac{1}{2} \,$} P+q)
       \Gamma(q,P)S(-\mbox{$ \, \frac{1}{2} \,$} P+q) 
       \gamma_\mu, \label{eq:BS}
\end{equation}
where $\Gamma(p,P)$ is the one fermion irreducible 
positronium-fermion-antifermion vertex with external legs 
amputated.  The photon 
propagator $D(p-q)$ in Feynman gauge is $1/(p-q)^2$.  The
fermion propagator $S$ is the solution to a truncated 
Schwinger-Dyson (SD) equation.  A fermion propagator has been chosen
which supports spontaneous mass generation necessary 
for the formation of the Goldstone bosons.  
The truncated SD equation for a fermion of bare mass $m$ is
\begin{equation}
\Sigma(p)= S(p)^{-1} - (i\!\not \! p + m) = 
 e^2 \int \mbox{$ \, \frac{d^3q}{(2\pi)^3} \,$} D(p-q) \gamma_\mu 
 S_F(q) \gamma_\mu.  \label{eq:SD}
\end{equation}
This approximation is the quenched, rainbow approximation
named so because the photon propagator has been replaced by the
bare photon propagator and the vertex function $\Gamma$ has been 
replaced by the bare vertex $\gamma$, resulting in a series
of Feynman diagrams which resemble rainbows.
In the quenched approximation the SD and BS equations can be 
recast in terms of a dimensionless momentum $p/e^2$ and bare fermion 
mass $m/e^2$.  From here on we work in dimensionless units and set 
$e^2 = 1$.  

We use either of the two following equivalent 
representations of the fermion propagator,
\begin{equation}
S(p)= -i \not\! p \sigma_V(p^2)+\sigma_S(p^2)
\label{eq:SIGDEF}
\end{equation}
or
\begin{equation}
S(p)= \frac{1}{ i \not\! p A(p^2)+B(p^2) }.   \label{eq:ABDEF}
\end{equation}
The generation of a dynamical fermion mass and the breaking of 
chiral symmetry is signalled (in the massless limit)  
by non-zero $B(p^2)$.  The vector and
scalar parts $\sigma_V$ and $\sigma_S$ of the propagator are
related to the functions $A$ and $B$ simply by dividing 
these functions ($A$, $B$)
by a quantity $p^2 A^2(p^2) + B^2(p^2)$.  

Note that a substitution of this fermion propagator into the 
Ward-Takahashi identity shows that the bare vertex approximation 
breaks gauge covariance.  However, this model is simple and does 
meet the requirement that the appropriate Goldstone
bosons are formed~\cite{DS79}.  
It is not difficult to derive a zero mass
solution to our BSE analytically. A vertex proportional to the
matrix $\gamma_4$ or $\gamma_5$, defined in the appendix, 
will reduce the quenched ladder 
BSE to the quenched ladder (rainbow) SDE in the case of zero bound
state mass thus forming a doublet of massless states.  According
to the terminology used in the appendix this is an axi-scalar
doublet.  These solutions will be seen in section 5.

Once the photon and fermion propagators are supplied, the BS
equation can be written as a set of numerically tractable
integral equations.  To do this, we write the BS
amplitude $\Gamma$ in its most general form consistent with
the parity and charge conjugation of the required bound
state, and then project out the coefficient functions for
the individual Dirac components.  
It is convenient to work
in the rest frame of the bound state by setting 
$P_{\mu}=(0,0,iM)$.  Then the scalar and axi-scalar vertices
given in the appendix by eqs.~(\ref{eq:GS}) and (\ref{eq:GAS})
can be written as
\begin{eqnarray}
\lefteqn{\Gamma^S(q,P) = 
f(q_3,\mbox{$\left| {\bf q} \right|$};M) - \frac{i q_j \gamma_j}
{\mbox{$\left| {\bf q} \right|$}} U(q_3,\mbox{$\left| {\bf q} 
\right|$};M) } \nonumber \\
& & + \frac{i q_j^{\perp} \gamma_j}{\mbox{$\left| {\bf q} 
\right|$}}\gamma_{45} 
V(q_3,\mbox{$\left| {\bf q} \right|$};M)  + i \gamma_3 
W(q_3,\mbox{$\left| {\bf q} \right|$};M), \label{eq:GSNEW}
\end{eqnarray}
\begin{eqnarray}
\Gamma^{AS}(q,P) & = & \left(\begin{array}{c} \gamma_4 \\
 \gamma_5\end{array} \right) f(q_3,\mbox{$\left| {\bf q} 
 \right|$};M) - \frac{i q_j \gamma_j}{\mbox{$\left| {\bf q} \right|$}} 
\left(\begin{array}{c} \gamma_4 \\ \gamma_5\end{array} \right) 
 U(q_3,\mbox{$\left| {\bf q} \right|$};M) \nonumber \\
 &  & + \frac{i q_j^{\perp} \gamma_j}{\mbox{$\left| {\bf q} \right|$}} 
\left(\begin{array}{c} \gamma_5 \\ -\gamma_4\end{array} \right) 
   V(q_3,\mbox{$\left| {\bf q} \right|$};M) - \gamma_3 \left(
   \begin{array}{c} \gamma_{4} \\ 
\gamma_{5}\end{array} \right) W(q_3,\mbox{$\left| {\bf q} 
\right|$};M), \label{eq:GASNEW}
\end{eqnarray}
where the index $j$ takes on values 1 and 2 only, 
$\mbox{$\left| {\bf q} \right|$}=(q_1^2+q_2^2)^{\frac{1}{2}}$ and 
$\mbox{$ \, {\bf q} \,$}^{\perp}=(-q_2,q_1)$.  
The pseudoscalar and axi-pseudoscalar vertices are obtained
from the scalar and axi-scalar vertices by multiplication by
the matrix $\gamma_{45}$.

It is found that the same coupled integral
equations result when a vertex is multiplied by the matrix
$\gamma_{45}$ and so (scalar, pseudoscalar) and (axi-scalar, 
axi-pseudoscalar) form two pairs of degenerate states.

The four equations derived from the BSE, after some manipulation
including an angular integration, are~\cite{Bu92}:
\begin{eqnarray}
f(p) & = & \frac{3}{(2\pi)^2} \int^{\infty}_{-\infty}dq_3 \,
     \int^{\infty}_0 \mbox{$\left| {\bf q} \right|$} d\mbox{$\left| 
     {\bf q} \right|$} \, \frac{1}{(\alpha^2-\beta^2)^{\frac{1}{2}}}
       \times\nonumber \\
 & &  (T_{ff}f(q)+T_{fU}U(q)+T_{fV}V(q)+T_{fW}W(q)) \nonumber \\
U(p) & = & \frac{1}{(2\pi)^2} \int^{\infty}_{-\infty}dq_3 \,
  \int^{\infty}_0 \mbox{$\left| {\bf q} \right|$} d\mbox{$\left| 
  {\bf q} \right|$} \, 
     \frac{(\alpha^2-\beta^2)^{\frac{1}{2}}-\alpha}
             {\beta(\alpha^2-\beta^2)^{\frac{1}{2}}}
       \times \nonumber \\
 & &   (T_{Uf}f(q)+T_{UU}U(q)+T_{UV}V(q)+T_{UW}W(q)) \nonumber \\
V(p) & = & \frac{1}{(2\pi)^2} \int^{\infty}_{-\infty}dq_3 \,
  \int^{\infty}_0 \mbox{$\left| {\bf q} \right|$} d\mbox{$\left| 
  {\bf q} \right|$} \, 
     \frac{(\alpha^2-\beta^2)^{\frac{1}{2}}-\alpha}
             {\beta(\alpha^2-\beta^2)^{\frac{1}{2}}}
       \times\nonumber \\
 & &   (T_{Vf}f(q)+T_{VU}U(q)+T_{VV}V(q)+T_{VW}W(q)) \nonumber \\
W(p) & = & \frac{1}{(2\pi)^2} \int^{\infty}_{-\infty}dq_3 \,
   \int^{\infty}_0 \mbox{$\left| {\bf q} \right|$} d\mbox{$\left| 
   {\bf q} \right|$} \, \frac{1}{(\alpha^2-\beta^2)^{\frac{1}{2}}}
       \times\nonumber \\
 & &  (T_{Wf}f(q)+T_{WU}U(q)+T_{WV}V(q)+T_{WW}W(q)) ,    \label{eq:IE}
\end{eqnarray}
where
$$
\alpha=(p_3-q_3)^2 + \mbox{$\left| {\bf p} \right|$}^2 + \mbox{
$\left| {\bf q} \right|$}^2, \;\;\;\;\;\;
\beta=-2 \mbox{$\left| {\bf p} \right|$} \mbox{$\left| {\bf q} \right|$}.
$$
Now define the momentum Q by
\begin{equation}
Q^2=q_3^2 + \mbox{$\left| {\bf q} \right|$}^2 -\frac{1}{4}M^2 
+ iMq_3 , \label{eq:QDEF}
\end{equation}
and use the abbreviations $\sigma_V=\sigma_V(Q^2)$ and 
$\sigma_S=\sigma_S(Q^2)$ for use in the definition of the 
functions $T_{ff},T_{fU},\ldots$ which are analytic 
functions of $q_3, \mbox{$\left| {\bf q} \right|$}$, and $M$.  
The diagonal $T$'s are given by 
\begin{eqnarray}
T_{ff}&=&(\frac{1}{4}M^2 + q_3^{\,2} + \mbox{$\left| {\bf q} 
\right|$}^2) |\sigma_V|^2 
  \mp |\sigma_S|^2, \nonumber \\
T_{UU}&=&(\frac{1}{4}M^2 + q_3^{\,2} - \mbox{$\left| {\bf q} 
\right|$}^2) |\sigma_V|^2 
  \pm |\sigma_S|^2, \nonumber \\
T_{VV}&=&(\frac{1}{4}M^2 + q_3^{\,2} + \mbox{$\left| {\bf q} 
\right|$}^2) |\sigma_V|^2 
  \pm |\sigma_S|^2, \nonumber \\
T_{WW}&=&-(\frac{1}{4}M^2 + q_3^{\,2} - \mbox{$\left| {\bf q} 
\right|$}^2) |\sigma_V|^2 
  \pm |\sigma_S|^2,
\end{eqnarray}
where the upper sign applies to the scalar equations
and the lower sign to the axi-scalar equations.
The off-diagonal $T$'s are, for the scalar positronium states:
\begin{equation}
\begin{array}{ccccl}
T_{fU}&=&T_{Uf}&=&(\sigma_V^{\ast}\sigma_S+\sigma_S^{\ast}\sigma_V)
\mbox{$\left| {\bf q} \right|$}, \\
T_{fV}&=&T_{Vf}&=&-M\mbox{$\left| {\bf q} \right|$}|\sigma_V|^2, \nonumber \\
T_{fW}&=&T_{Wf}&=&-(\sigma_V^{\ast}\sigma_S+\sigma_S^{\ast}\sigma_V)q_3
           +\frac{i}{2}(\sigma_V^{\ast}\sigma_S-\sigma_S^{\ast}\sigma_V)M,  \\
T_{UV}&=&T_{VU}&=&-[\frac{1}{2}(\sigma_V^{\ast}\sigma_S+
                                     \sigma_S^{\ast}\sigma_V)M
                 +iq_3(\sigma_V^{\ast}\sigma_S-\sigma_S^{\ast}\sigma_V)],  \\
T_{UW}&=&T_{WU}&=&2q_3\mbox{$\left| {\bf q} \right|$}|\sigma_V|^2,  \\
T_{VW}&=&T_{WV}&=&-i(\sigma_V^{\ast}\sigma_S-\sigma_S^{\ast}\sigma_V)
\mbox{$\left| {\bf q} \right|$},
\end{array}
\end{equation}
and for the axi-scalar states:
\begin{equation}
\begin{array}{ccccl}
T_{fU}&=&T_{Uf}&=&i(\sigma_V^{\ast}\sigma_S-\sigma_S^{\ast}\sigma_V)
\mbox{$\left| {\bf q} \right|$},  \\
T_{fV}&=&-T_{Vf}&=&M\mbox{$\left| {\bf q} \right|$}|\sigma_V|^2,  \\
T_{fW}&=&-T_{Wf}&=&-i(\sigma_V^{\ast}\sigma_S-\sigma_S^{\ast}\sigma_V)q_3
          -\frac{1}{2}(\sigma_V^{\ast}\sigma_S+\sigma_S^{\ast}\sigma_V)M,  \\
T_{UV}&=&T_{VU}&=&-
       [\frac{i}{2}(\sigma_V^{\ast}\sigma_S-\sigma_S^{\ast}\sigma_V)M
                 -q_3(\sigma_V^{\ast}\sigma_S+\sigma_S^{\ast}\sigma_V)],  \\
T_{UW}&=&-T_{WU}&=&-2q_3\mbox{$\left| {\bf q} \right|$}|\sigma_V|^2, \\
T_{VW}&=&-T_{WV}&=&(\sigma_V^{\ast}\sigma_S+\sigma_S^{\ast}\sigma_V)
\mbox{$\left| {\bf q} \right|$}.
\end{array}
\end{equation}
This is the same set of equations solved in Ref.~\cite{Bu92} 
with only the fermion propagator input altered.  The bare fermion 
mass $m$ only comes into the calculation through this input.

The solution to the BSE involves iteration of the coupled integral 
equations in Eq.~(\ref{eq:IE}).  These equations may be rewritten as
\begin{equation}
{\bf f}(\mbox{$\left| {\bf p} \right|$},p_3;M) = \int dq_3
 \int d\mbox{$\left| {\bf q} \right|$} \, K(\mbox{$\left| {\bf p} 
 \right|$},p_3;\mbox{$\left| {\bf q} \right|$},q_3;M) {\bf f}
 (\mbox{$\left| {\bf q} \right|$},q_3;M),
                      \label{eq:BSI}
\end{equation}
where ${\bf f}=(f,U,V,W)^{\rm T}$. For each symmetry case and each 
fermion mass this is solved as an eigenvalue problem of the form
\begin{equation}
\int dq\,K(p,q;M) {\bf f}(q) = \Lambda(M) {\bf f}(p), \label{eq:EV}
\end{equation}
for a given test mass $M$.  This is repeated for different test 
bound state masses until an eigenvalue $\Lambda(M)=1$ is obtained.

\setcounter{equation}{0}
\subsection{Non-Relativistic Limit}
We consider now the non-relativistic limit $m\rightarrow \infty$ 
of our BS formalism in order to enable comparisons with existing 
numerical calculations \cite{YH91,THY95} of the Schr\"odinger 
equation for QED3, and with the large $m$ solution of 
Eq.~(\ref{eq:IE})

The Schr\"odinger equation with a confining logarithmic potential 
is an interesting problem in its own right.  Initially 
one is faced with the problem of setting the scale of the 
potential, or equivalently, setting the zero of energy of the 
confined bound states.  
A solution to this problem was proposed by Sen~\cite{S90} and 
Cornwall~\cite{C80} in terms of cancellation of infrared 
divergences in perturbation theory.  They introduce a regulating 
photon mass $\mu$ in order to set 
the potential as the 2-dimensional Fourier transform of the photon 
propagator $1/(k^2 + \mu^2)$, leading to a potential proportional to 
$\ln (\mu r)$.  They further interpret the sum of the bare fermion mass and the fermion self energy evaluated at the bare fermion 
mass shell as a renormalised fermion mass, leading to a mass 
renormalisation $\delta m \propto \ln (m/\mu)$.  The logarithmic 
divergences in the photon potential and the fermion self energy 
then conspire to cancel leaving a finite positronium mass.  

The first numerical treatment of the Schr\"odinger 
equation for QED3 using this line of argument was carried out by 
Yung and Hamer \cite{YH91}. In a subsequent, improved calculation by 
Tam, Hamer and Yung \cite{THY95}, the formalism was shown to 
be consistent with an analysis of QED3 from the point of view of discrete 
light cone quantisation.  Their resulting expression for the bound 
state energy, obtained as a solution to the differential equation 
\begin{equation} 
\left\{ -\frac{1}{m} \nabla^2 + \frac{1}{2\pi} \left(C + \ln(mr) 
 \right)\right\}  \phi({\bf r}) = (E - 2m) \phi({\bf r}), \label{eq:THYDE}     
\end{equation}
where $C$ is Euler's constant, is  
given in terms of the bare fermion mass $m$ as 
\begin{equation}
E  =  2m + \frac{1}{4\pi} \ln m  + \frac{1}{2\pi} \left(\lambda -
\frac{1}{2} \ln\frac{2}{\pi} \right).  
       \label{eq:THY1}
\end{equation}
The lightest s-wave positronium state and first exited state are given by 
$\lambda_0 = 1.7969$ and $\lambda_1 = 2.9316$ respectively.  The first five 
states are provided in ref.~\cite{THY95}.  

Here we present a treatment of the non-relativistic limit of the 
QED3 positronium spectrum in terms of our SD--BS equation formalism.  
We begin with the fermion propagator in the limit 
$m\rightarrow \infty$.  
For large fermion mass we expect the
residual effect of the chiral symmetry breaking contribution
to the fermion self energy to be small compared with contribution from the
perturbative loop expansion.  We shall therefore assume to begin with that 
the self energy is reasonably well approximated by the one-loop result.  
The validity of this approximation for spacelike momenta will be 
demonstrated  numerically in the next section.  

The 1-loop fermion self energy, with the functions $A$ and $B$ 
defined in Eq.~(\ref{eq:ABDEF}), is given by 
\begin{equation}
A = 1 + \frac{\Sigma_A(\frac{p^2}{m^2})}{m},
\label{eq:A1LOOP}
\end{equation}
where
\begin{equation}
\Sigma_A(x^2)=\frac{1}{8\pi x^2} \left[ 1 - \frac{1-x^2}{2x} 
   \arccos\left( \frac{1-x^2}{1+x^2} \right) \right],
\label{eq:SIGA1}
\end{equation}
and
\begin{equation}
B = m \left( 1 + \frac{\Sigma_B(\frac{p^2}{m^2})}{m} \right),
\label{eq:B1LOOP}
\end{equation}
where
\begin{equation}
\Sigma_B(x^2)=\frac{3}{8\pi x} \arccos\left( \frac{1-x^2}{1+x^2} \right) .
\label{eq:SIGB1}
\end{equation}
This result is valid for (Euclidean) spacelike momenta $p^2 > 0$.  An analytic 
continuation of the $\Sigma$ functions valid for $|p|^2 < m^2$, or 
$\left| x \right| <1$, is
\begin{equation}
\Sigma_A(x^2)=\frac{1}{8\pi i x} \left[ \frac{x^2-1}{2x^2} 
   \ln\left( \frac{1+ix}{1-ix} \right) + \frac{i}{x} \right],
\end{equation}
\begin{equation}
\Sigma_B(x^2)=\frac{3}{8\pi i x} \ln\left( \frac{1+ix}{1-ix} \right) .
\end{equation}
Note that this representation exposes a logarithmic infinity in the 
self energy at the bare fermion mass pole $p^2=-m^2$.  
This is the 
infrared divergence in the renormalised fermion self energy as defined by 
Sen~\cite{S90} referred to above.  However, in our formalism, this 
singularity does not lead to a pole in the propagator functions 
$\sigma_V$ and $\sigma_S$ defined in Eq.~(\ref{eq:SIGDEF}),
which would signal the propagation of a free fermion~\cite{RW94},
but a logarithmic zero.  

Using Eqs.~(\ref{eq:A1LOOP}) and (\ref{eq:B1LOOP}) we obtain 
\begin{equation}
\sigma_V(p^2) = \frac{1}{m^2}\frac{1 + \frac{\Sigma_A}{m}}
     {\epsilon\left(1 + \frac{\Sigma_A}{m}\right)^2 + 
     2\left(1 + \frac{\Sigma_A + \Sigma_B}{2m}\right) 
                         \frac{\Sigma_B - \Sigma_A}{m}}, 
\end{equation}
\begin{equation}
\sigma_S(p^2) = \frac{1}{m}\frac{1 + \frac{\Sigma_B}{m}}
     {\epsilon\left(1 + \frac{\Sigma_A}{m}\right)^2 + 
     2\left(1 + \frac{\Sigma_A + \Sigma_B}{2m}\right) 
                         \frac{\Sigma_B - \Sigma_A}{m}}, 
\end{equation}
where we have defined 
\begin{equation}
\epsilon = \frac{p^2 + m^2}{m^2}.
\end{equation}  
The functions $\sigma_V$ and $\sigma_S$
are plotted in Figs.~2a and 2b for $m=$ 1, 2, 4, 8 and $\infty$,
the final curve being the bare propagator.  From these plots we see that 
for large $m$, the deviation from the bare propagator due to the 1-loop 
self energy is dominated by the logarithmic contribution near the bare 
fermion mass shell, $\epsilon = 0$.  With this in mind, we shall use the 
approximation 
\begin{equation}
\Sigma_A(x^2) \approx -\frac{1}{8\pi} \ln\frac{\epsilon}{4},
\label{eq:SIGAP1}
\end{equation}
\begin{equation}
\Sigma_B(x^2) \approx -\frac{3}{8\pi} \ln\frac{\epsilon}{4}.
\label{eq:SIGAPX}
\end{equation}
Taking $\epsilon$ to be of order $1/m$ for the purposes of the BS equation 
(see Eq.~(\ref{eq:EPSAPX}) below) these approximations give 
\begin{equation}
S(p) = \frac{-i \! \not \! p + m}{m} \cdot \frac{1}{m\epsilon
                  - \frac{1}{2\pi} \ln \frac{\epsilon}{4}} 
         \left(1 + O\left(\frac{\ln m}{m}\right)\right).  \label{eq:SAPX}
\end{equation}
The vector and scalar parts of this approximate propagator (without 
the $O(\ln m /m)$ corrections) are also plotted in Figs.~2a and 2b for 
comparison.   

Turning now to the BS equation, we set the bound state momentum
in Eq.~(\ref{eq:BS}) equal to
$P_{\mu}=(2m+\delta)iv_{\mu}$, where $v_{\mu}=(0,0,1)$ and
$-\delta$ is a ``binding energy''.  This gives (according to the
momentum distribution in Fig.~1)
\begin{equation}
\Gamma(p) = - \int \frac{d^3q}{(2\pi)^3} D(p-q) \, \gamma_\mu \, 
S\left[-(m+\frac{\delta}{2})iv_\mu+q_\mu\right]
       \Gamma(q) \, S\left[(m+\frac{\delta}{2})iv_\mu+q_\mu\right] \gamma_\mu. 
\label{eq:NRBSE}
\end{equation}
Setting 
\begin{equation}
\epsilon \, = \, \frac{2}{m}\left(-\frac{1}{2}\delta+iq_3+
       \frac{\left|{\bf q}\right|^2}{2m}\right) \, + \, 
O\left(\frac{1}{m^2}\right)
\label{eq:EPSAPX}
\end{equation}
in Eq.~(\ref{eq:SAPX}) gives
\begin{equation}
S\left[(m+\frac{\delta}{2})iv_\mu+q_\mu \right]
 = \frac{1+\gamma_3}{2}\, \frac{1}{(\mbox{$ \, 
 -\frac{1}{2} \,$} \delta+iq_3+\frac{\left|{\bf q}\right|^2}{2m})-
\frac{1}{4\pi} \ln \frac{\epsilon}{4}} \, + \, O\left(\frac{\ln m}{m}\right).
\label{eq:SPLUS}
\end{equation}
Similarly
\begin{equation}
S\left[-(m+\frac{\delta}{2})iv_\mu+q_\mu \right]
 = \frac{1-\gamma_3}{2}\, \frac{1}{(\mbox{$ \, -\frac{1}{2} \,$} 
 \delta-iq_3+\frac{\left|{\bf q}\right|^2}{2m})-
\frac{1}{4\pi} \ln \frac{\epsilon^*}{4}} \, + \, O\left(\frac{\ln m}{m}\right).
\label{eq:SMINUS} 
\end{equation}
The $\left|{\bf q}\right|^2/2m$ term has been retained here to ensure 
convergence of the $\left|{\bf q}\right|$ integral in the BS equation below.  
Since the vertex $\Gamma$ is defined with the fermion legs 
truncated, and $S \propto \frac{1}{2}(1 \pm \gamma_3)$, the only
relevant part of $\Gamma$ is the projection 
$\mbox{$ \, \frac{1}{2} \,$}(1-\gamma_3) \, \Gamma \, 
\mbox{$ \, \frac{1}{2} \,$}(1+\gamma_3)$.  
With this in mind,
the general forms in eqs.~(\ref{eq:GSNEW}) and (\ref{eq:GASNEW}) 
become,
$$
\mbox{$ \, \frac{1}{2} \,$}(1-\gamma_3) \, \Gamma^S \, 
\mbox{$ \, \frac{1}{2} \,$}(1+\gamma_3)  =  
\mbox{$ \, \frac{1}{2} \,$}(1-\gamma_3) \frac{q_j 
\gamma_j}{\mbox{$\left| {\bf q} \right|$}}  g(q_3,\mbox{$\left| 
{\bf q} \right|$})
$$
\begin{equation}
\mbox{$ \, \frac{1}{2} \,$}(1-\gamma_3) \, \Gamma^{AS} \, 
\mbox{$ \, \frac{1}{2} \,$}(1+\gamma_3)  =  
\mbox{$ \, \frac{1}{2} \,$}(1-\gamma_3) \left(\begin{array}{c} 
\gamma_4 \\
 \gamma_5\end{array} \right)  g(q_3,\mbox{$\left| {\bf q} \right|$}).
\label{eq:SASEQ}
\end{equation}

Substituting eqs.~(\ref{eq:SPLUS}), (\ref{eq:SMINUS}) and
(\ref{eq:SASEQ}) into Eq.~(\ref{eq:NRBSE}) one obtains for 
the scalar states the single integral equation
\begin{equation}
g(p) \approx\int \frac{d^3q}{(2\pi)^3} \,
     \frac{1}{(p - q)^2} 
     \frac{\bf{p.q}}{\left|{\bf p}\right|\left|{\bf q}\right|}
     \frac{g(q)}{ \left|-\frac{1}{2}\delta+
     iq_3 + \frac{\left|{\bf q}\right|^2}{2m} - 
     \frac{1}{4\pi} \ln \frac{1}{2m}
     (-\frac{1}{2}\delta+iq_3+\frac{\left|{\bf q}\right|^2}{2m})
                        \right|^2 },
\label{eq:NREQ1}
\end{equation}
and for the axi-scalar states the single equation 
\begin{equation}
g(p) \approx \int \frac{d^3q}{(2\pi)^3}\,
     \frac{1}{(p - q)^2} 
     \frac{g(q)}{ \left|-\frac{1}{2}\delta+
     iq_3 + \frac{\left|{\bf q}\right|^2}{2m} - 
     \frac{1}{4\pi} \ln \frac{1}{2m}
     (-\frac{1}{2}\delta+iq_3+\frac{\left|{\bf q}\right|^2}{2m})
                      \right|^2 }.
\label{eq:NREQ1A}
\end{equation}
Note that, without the $\left|{\bf q}\right|^2/2m$ term in the 
denominator, translation 
invariance of the integrand implies that $g$ is independent of 
$\left|{\bf q}\right|$.  In reality, $g$ is a slowly varying function of  
$\left|{\bf q}\right|$, and this extra $O(1/m^2)$ term must be retained 
in $\epsilon$ to account for the fact that the relevant region of 
integration in Eqs.~(\ref{eq:NREQ1}) and (\ref{eq:NREQ1A}) extends out to 
$O(\sqrt{m})$ in the $\left|{\bf q}\right|$ direction, but only $O(1)$ in 
the $q_3$ direction.  Numerical solutions of Eqs.~(\ref{eq:NREQ1}) and 
(\ref{eq:NREQ1A}) will be given in section 5.
The function $g$ is an even or odd function of $q_3$ corresponding 
to positronium states which are even or odd respectively under 
charge conjugation.    

In order to obtain a Schr\"{o}dinger equation, we now rewrite the 
axi-scalar equation in the form 
\begin{eqnarray}
g(p) & = & \int \frac{d^2{\bf }q}{(2\pi)^2} \int_{-\infty}^\infty 
    \frac{dq_3}{2\pi}\,
     \frac{1}{(p_3 - q_3)^2 + \left|{\bf p - q}\right|^2} 
     \times \nonumber \\
   & &   \!\!\!\!\! \frac{g(q)}{ \left(-\frac{1}{2}\delta+
     iq_3 + \frac{\left|{\bf q}\right|^2}{2m} 
     + \Sigma_+(q_3,\left|{\bf q}\right|)\right)
     \left(-\frac{1}{2}\delta -
     iq_3 + \frac{\left|{\bf q}\right|^2}{2m} 
     + \Sigma_-(q_3,\left|{\bf q}\right|)\right)},
\label{eq:NREQ2}
\end{eqnarray}
where
\begin{equation}
\Sigma_\pm(q_3,\left|{\bf q}\right|) = 
     - \frac{1}{4\pi} \ln \frac{1}{2m}
     \left(-\frac{1}{2}\delta \pm iq_3
           +\frac{\left|{\bf q}\right|^2}{2m}\right).  \label{eq:SIGPM}
\end{equation}
Assuming the integrand dies off sufficiently rapidly as 
$q_3 \rightarrow -i\infty$, we deform the contour of integration around 
the pole at 
\begin{equation}
q_3^{\rm pole} = -i \left(-\frac{1}{2}\delta + 
          \frac{\left|{\bf q}\right|^2}{2m} 
              + \Sigma_-(q_3^{\rm pole},\left|{\bf q}\right|) \right),
                   \label{eq:Q3POLE}
\end{equation}
to obtain 
\begin{equation}
g(p) = \int \frac{d^2{\bf q}}{(2\pi)^2} \,
     \frac{1}{(p_3 - q_3^{\rm pole})^2 + \left|{\bf p - q}\right|^2} 
     .\frac{g(q_3^{\rm pole},\left|{\bf q}\right|)}
      {-\delta + \frac{\left|{\bf q}\right|^2}{m} 
     + 2\Re\Sigma_-(q_3^{\rm pole},\left|{\bf q}\right|)}.  
\label{eq:NREQ3}
\end{equation}
(We could equally well deform the contour round the pole at 
$(q_3^{\rm pole})^*$ if the integrand decays in the opposite direction, 
without affecting our final result.)  

Defining 
\begin{equation} 
\Phi(p_3,\left|{\bf p}\right|) =  \frac{g(p_3,\left|{\bf p}\right|)}
      {-\delta + \frac{\left|{\bf p}\right|^2}{m} 
     + 2\Re\Sigma_-(p_3,\left|{\bf p}\right|)}, 
\end{equation} 
gives   
\begin{eqnarray} 
\lefteqn{\left\{-\delta + \frac{\left|{\bf p}\right|^2}{m} 
     + 2\Re\Sigma_-(p_3,\left|{\bf p}\right|)\right\}
        \Phi(p_3,\left|{\bf p}\right|) } \hspace{30 mm}\nonumber \\
 & = & \int \frac{d^2{\bf q}}{(2\pi)^2} \, 
     \frac{1}{(p_3 - q_3^{\rm pole})^2 + \left|{\bf p - q}\right|^2} 
        \Phi(q_3^{\bf pole},\left|{\bf q}\right|). \label{eq:NRPHI1}
\end{eqnarray}
In order to isolate the logarthmic infrared divergence we set 
\begin{equation}
p_3 = p_3^{\rm pole} + \mu,   \label{eq:P3MU} 
\end{equation}
with $\mu$ small and real, and $p_3^{\rm pole}$ defined by analogy with 
Eq.~(\ref{eq:Q3POLE}).  The right hand side of Eq.~(\ref{eq:NRPHI1}) 
then becomes 
\begin{eqnarray}
\mbox{r.h.s.} & = & \int \frac{d^2{\bf q}}{(2\pi)^2} \, 
     \frac{1}{\mu^2 + O(\mu(\left|{\bf p}\right| - \left|{\bf q}\right|))
              + \left|{\bf p - q}\right|^2} \phi({\bf q}) \nonumber \\
 & = & \mbox{F.T. of } \frac{-1}{2\pi} \left[C + 
        \ln\left(\frac{\mu r}{2}\right)\right] \phi({\bf r})
         \hspace{5 mm} \mbox{as $\mu \rightarrow 0$}, \label{eq:RHS}
\end{eqnarray}
where $\phi({\bf q}) = \Phi(q_3^{\bf pole},\left|{\bf q}\right|)$.  

Following the reasoning of refs.~\cite{S90} and \cite{C80}, this 
logarithmic divergence should be cancelled by the fermion self energy 
contribution $2\Re\Sigma_-(p_3,\left|{\bf p}\right|)$.  However, 
from Eq.~(\ref{eq:SIGPM}), we see that the logarithmic divergence in 
the self energy occurs at the bare fermion mass pole 
$p_3^{\rm bare} = -i (-\delta/2 + \left|{\bf p}\right|^2/2m)$, and not 
the dressed pole $p_3^{\bf pole}$.  The problem lies in the use of the 
1-loop approximation.  If instead the fermion self energy is calculated 
to all orders in rainbow approximation, the self energy feeds back into 
the loop integral via the propagator to replace Eq.~(\ref{eq:SIGPM}) by 
\begin{equation}
\Sigma_-(p_3,\left|{\bf p}\right|) = 
     - \frac{1}{4\pi} \ln \frac{1}{2m}
     \left(-\frac{1}{2}\delta - ip_3
           +\frac{\left|{\bf p}\right|^2}{2m} + 
  \Sigma_-(p_3,\left|{\bf p}\right|)\right),  \label{eq:SIGM}
\end{equation}
which provides a rainbow SD equation for $\Sigma_-$ 
in the non-relativistic limit.  
Then using Eqs.~(\ref{eq:NRPHI1}), (\ref{eq:Q3POLE}), (\ref{eq:P3MU}), 
(\ref{eq:RHS}) and (\ref{eq:SIGM}) and fourier transforming we finally 
obtain 
\begin{equation} 
\left\{ -\frac{1}{m} \nabla^2 + \frac{1}{2\pi} \left(C + \ln(mr) 
 \right)\right\}  \phi({\bf r}) = \delta \phi({\bf r}),     
\end{equation}
agreeing with Eq.~(\ref{eq:THYDE}). 
Had we started from the scalar equation (\ref{eq:NREQ1}) in place of the 
axi-scalar equation, the same result would have been obtained at 
Eq.~(\ref{eq:RHS}), leading to an identical Schr\"{o}dinger equation.  

The important point to notice in this derivation is the significance 
of a non-perturbative solution to the SD equation in cancelling the 
infrared divergences.  In the massless fermion limit, it is well known 
that chiral symmetry breaking plays a pivotal role in determining the 
bound state spectrum.  It appears also that, even in the 
non-relativistic limit, the remnant effects of chiral symmetry 
breaking, via a non-perturbative solution to the SD equation, have a 
role to play.  

\setcounter{equation}{0}
\subsection{The Fermion Propagator}
The BSE described in Section 2 requires a fermion
propagator input in the form of Eq.~(\ref{eq:SIGDEF}) or 
Eq.~(\ref{eq:ABDEF}) and this
needs to be available over a region in the complex $p^2$ plane
defined by Eq.~(\ref{eq:QDEF}) for $q_3$ and $\mbox{$\left| {\bf q} 
\right|$}$ real.  This
is the region~\cite{SC92,SC94}
\begin{equation}
\Omega = \left\{ Q^2 = X + iY \left| X > \frac{Y^2}{M^2} 
- \frac{1}{4} M^2 \right. \right\}.
\label{eq:REGION}
\end{equation}
In this section we investigate ways of obtaining a solution to the
SDE over $\Omega$.  The fermion propagator, and thus the functions
$\sigma_V$ and $\sigma_S$, must be well behaved over this region.

The solution to the SDE~(\ref{eq:SD}) is quite simple
along the positive real (spacelike)-$p^2$ axis. Substitution of the
general expression for the fermion propagator~(\ref{eq:ABDEF})
into the SDE
gives an integral equation involving $A$ and $B$ functions
which can be split into two coupled integral equations by
simple projections.  Angular integrals can be performed to
leave one dimensional integrals (over the modulus of the $q$ 
vector),
$$
A(p^2) - 1 = \frac{1}{4\pi^2p^2} \int^{\infty}_{0}dq 
     \frac{q A(q^2)}{q^2 A^2(q^2)+B^2(q^2)} 
     \left({\frac{p^2+q^2}{4 p} 
     \ln\left({\frac{p+q}{p-q}}\right)^2 - q}\right), 
$$

\begin{equation}
B(p^2) - m = \frac{3}{8\pi^2p} \int^{\infty}_{0}dq 
     \frac{q B(q^2)}{q^2 A^2(q^2)+B^2(q^2)} 
     \ln\left({\frac{p+q}{p-q}}\right)^2  
      \label{eq:ABINTS}
\end{equation}
The integrations range from 0 to some UV cutoff along
the positive real axis.  This theory is
super-renormalisable and thus has no ultraviolet divergences and
so this cutoff is merely a numerical limit made large enough so
that it has no bearing on the results.

For a set of points $p$ corresponding
to the set of $q$ points in the integration, the equations 
are iterated until convergence to leave the solution along the
positive real axis.  However, the solution is required for complex
$p^2$.  We see three possibilities. The first is to use the
converged functions $A(q^2)$ and $B(q^2)$ in the integrals over the 
same contour (positive real $q^2$) and
supply the complex point $p$ desired.  The integrals should
provide the solution at that point $p$.  However the analytic structure 
of the integrands in Eq.~(\ref{eq:ABINTS}) will not allow an analytic 
continuation by this method, because a pinch singularity in the 
integrand forces us to integrate through the point $p$~\cite{Ma95}.  

The second possibility is to rotate the contour through an angle
$2\phi$ in the $p^2$ plane so that it passes 
through the desired point $p$~\cite{SC92}. In this way a cancellation of the 
complex parts within the logarithms occurs.  Fig.~3 shows the
first and second contours ($C_1$ and $C_2$ respectively).    
It can be seen from Eq.~(\ref{eq:ABINTS}) that the
logarithms will have real arguments along the radial portion of $C_2$, while 
the arc portion contributes nothing to the integral because the integrand 
falls off sufficiently quickly in the ultraviolet~\cite{Ma95}.

Based on the Landau gauge calculations of Maris~\cite{Ma93,Ma95} we 
expect conjugate singularities to occur 
in the second and third quadrants of the $p^2$ plane away from 
the negative real (timelike) axis. Thus, as $2\phi$ increases towards $\pi$ 
from zero (and the negative real $p^2$ axis is approached)
a singularity interferes and we may have
convergence problems.
It will be seen that these singularities can lie a fair way
from $\phi=\frac{\pi}{2}$ and 
convergence problems can occur for $\phi$ not much more
than $\frac{\pi}{4}$ (ie: barely reaching into the second 
quadrant of $p^2$).
For the case $m=0$ to be considered shortly, 
no solution could be found for $\phi$ greater than 
0.90 radians (with a reasonable convergence criterion 
$\frac{\Delta B}{B}<0.001$).
For this solution to be applied to the BSE we need to know 
the value of the fermion
propagator for $\phi$ from $0$ to $\frac{\pi}{2}$ and so this
method is not practical.
However, although a slowing of convergence as $\phi$ increases prevents
a solution being attained in all of $\Omega$, it does provide an
accurate solution in a large portion of $\Omega$.
We therefore
have a test for any $A$ and $B$ functions we wish to use in the 
BSE.  

The third possibility, and the one employed here and in ref.~\cite{Bu92},
is to find a good analytic fit along the positive real $p^2$ 
axis and extend the 
solution into the complex plane by analytic continuation.
These fits may be for $A$ and $B$ or for the functions
$\sigma_V$ and $\sigma_S$.  The work of Maris~\cite{Ma95} suggests
that it is not necessary for $\sigma_V$ and $\sigma_S$ to be
entire functions for the fermions to be confined, only that there 
be no poles on the timelike $p^2$ axis.
Fits to functions $A$ and $B$ used in previous work~\cite{Bu92} 
based on the known asymptotic infrared and ultraviolet
behaviour of these functions were
tested by comparing them with the direct solution for 
various angles $\phi$.
The fits, adjusted to allow variable fermion mass $m$, are given by 
$$
A_{\rm fit}(p^2)= \frac{a_1}{(a_2^{\,2} +p^2)^{\frac{1}{2}}}
             +a_3 e^{-a_4p^2} +1,
$$
\begin{equation}
B_{\rm fit}(p^2)= \frac{b_1}{b_2 +p^2}+b_3 e^{-b_4p^2} + m.  \label{eq:FIT}
\end{equation}
The parameters $a_n$,$b_n$ are functions of fermion mass.  
The numerical solution to which these functions were 
fitted is an
iterative solution to the SDE using a non-uniform 51 point grid
along the positive real axis up to a momentum cutoff 
$p = 1000$ using a $0.1\%$ tolerance in the 
integration routine.
Plots of the numerical solutions and function fits for various 
$m$ values are given in Fig.~4.

Note that it is the $\sigma$ functions that are important in
Eq.~(\ref{eq:BS}) and
not $A$ and $B$, and thus the effect of the fit on the denominator 
$p^2 A^2 + B^2$ relating these must be considered.
Conjugate poles exist where the factor $p^2 A^2 + B^2$ 
appearing in the denominator of the BSE integrand is zero.  
Table 1 lists the conjugate poles arising from the fits for each fermion mass
and the corresponding maximum bound state masses allowed.  The
maximum $M$ allowed is the value for which the boundary of 
$\Omega$ in Eq.~(\ref{eq:REGION}) coincides with the
conjugate poles. 
No comment about the viability of our model BSE can be made
until solutions are attempted because the integration
region depends on the solution mass $M$.

The location of the conjugate singularities for the $m=0$
case in Table 1 is slightly different to that reported in 
Ref.~\cite{Bu92} where it is $-0.00400 \pm i0.00666$.
This is because of the flexibility of the fitting functions.
The fit in this work and that in Ref.~\cite{Bu92} for the zero
fermion mass case had similar accuracy along the positive real
$p^2$ axis but had the freedom to take on slightly different
forms throughout the complex plane.  This is because along the 
positive real $p^2$ axis the non-asymptotic form fixing
parameters ($a_4$ and $b_4$) are only
loosely determined.  Despite the difference in the two
results, the BSE calculation for bound state masses should show close 
agreement as each fit adequately models the direct solution 
throughout the complex plane.

The singularities in the $\sigma$ fits
for fermion masses greater
than or equal to $0.1$ lie on the negative real $p^2$-axis.  
This suggests
that free propagation occurs at these masses and the model
is not confining.  An accurate location of the singularities in
the SDE solution would be needed before it can be said
whether this result is due to the fits or the rainbow 
approximation used in the SDE solution.
According to Ref.~\cite{Ma95} the rainbow approximation
SDE solution is expected to be confining even for 
large fermion mass.  Thus we assume our result is due to the
lack of accuracy in our fits near the negative real 
$p^2$ axis
and that it is likely that the singularities move close
to that axis as $m$ increases but never actually lie on that
axis.

Figs.~5a and 5b show plots of $\sigma_V$ and $\sigma_S$ moduli
respectively for zero fermion mass and angles $\phi=0$, 
$\phi=\frac{\pi}{8}$ and $\phi=\frac{\pi}{4}$ against the
$p$ modulus (with a range far smaller than the UV cutoff 
used in our calculations).  
The direct solutions to the SDE and the fits are compared.  It can 
be seen that the functions are very good fits along the positive 
real axis ($\phi=0$), where both $\sigma_V$ and $\sigma_S$ are real.
The fit is also good for $\phi=\frac{\pi}{8}$.  Real and imaginary 
components have not been given separately as they show similar 
agreement.  In the case $\phi=\frac{\pi}{4}$ the fitting function 
has begun to deviate from the SDE solution.  This is mostly due
to the apparent difference in the location of a spike.
Based on the largest bound state mass for $m=0$ reported 
in the next section, the BSE integration region $\Omega$ extends 
along the 
direction $\phi=\frac{\pi}{4}$ out to a modulus of approximately $0.083$.  
In this range the small angle solutions are very accurate but 
for larger $\phi$, much of the error due to the difference 
in the location
of the spike will be experienced.  
As the angle is increased further
convergence problems occur until eventually no solution can be 
found at all ($\phi>0.90$).    

The spike forming in these plots signals that, as $\phi$
is increased, the contour of integration approaches a singularity.
In fact, the conjugate poles which lie just off (or on as is the 
case for larger $m$) the negative real $p^2$ axis 
($\phi=\frac{\pi}{2}$) are approached.
It is important that both the direct solution
to the SDE and the fits used in this work have this feature.  
This spike was not seen in any other fits which we attempted.  
Based on Fig.~5 it seems clear that 
the direct solution to the SDE must have singularities 
close to those in the fitting functions.    

Because the spikes are not in 
exactly the same places some error will be introduced 
in the contributions from the large $\phi$ part of $\Omega$.  
When the bound state
mass becomes large, the large $\phi$ contributions will become
more important and thus we expect the error in the position of 
the spikes to result in some noise in the solutions to the 
BS equation for large fermion mass.  

The $\sigma$ functions were studied for all fermion masses used
in this work in the same fashion.  The results were similar 
to the $m=0$ case and need not be shown here.  In each case, 
when $\phi$ was increased
far enough, a spike was observed in both the fit and solution, 
after which lack of convergence prevented an SDE solution.

However, for very large fermion masses, the accuracy of the fits 
decreases as $m$ increases, and with good reason.  
As $m$ tends to infinity, the functions 
$A$ and $B$ approach constants ($1$ and $m$ respectively).  For
moderately large fermion masses experienced in this work, these
functions become almost constant along the positive real $p^2$
axis while having a singularity near the negative real $p^2$-axis.
It is too much to ask for simple four parameter fits along
the positive real $p^2$-axis to reproduce accurately complex behaviour
deep into the real timelike $p^2$-axis.  The 1-loop propagators, 
Fig.~2 described in section 3 illustrate this well.  There
one can see how smooth and level the $\sigma$ functions are
along the positive real $p^2$ axis and also how steep the functions
become back along the negative real $p^2$ axis.

Before moving on to the next section, we return briefly to the 
1-loop approximation to the fermion propagator necessary
for the non-relativistic approximations described in section 3.
Fig.~6a compares our rainbow approximation solution $A$ 
to the 1-loop result given in Eqs.~(\ref{eq:A1LOOP})
and (\ref{eq:SIGA1}).
Fig.~6b compares $B$ from our rainbow approximation
solution and the result in Eqs.~(\ref{eq:B1LOOP}) and
(\ref{eq:SIGB1}).
Both of these comparisons were made at a large fermion mass
($m=5$).  It can be seen
that the curves in each case are in reasonable agreement, at least 
for spacelike momenta.  

\setcounter{equation}{0}
\subsection{Numerical Solution of the Bethe-Salpeter Equation}
The fits given by Eq.~(\ref{eq:FIT}) to the fermion propagator for a range 
of fermion masses were used in the solution of the 
Bethe-Salpeter coupled integral equations Eq.~(\ref{eq:IE}).  This
problem was restated in Eq.~(\ref{eq:EV}) as an eigenvalue problem. 

A grid of $25 \times 25$ ($\mbox{$\left| {\bf q} \right|$}$,$q_3$) 
tiles were used for the 
iterative procedure with linear interpolation on each of those 
tiles used for the sums ($T_{ij}f_j$) which are 
supplied at the corners of the tiles from the previous 
iteration.  The tiles were non-uniform in size and an upper limit
to the momentum components ($\mbox{$\left| {\bf q} \right|$}$ and $q_3$) of 
between $3.0$ and $9.0$ was
used.  The equations were 
iterated to convergence each time
to determine eigenvalues for a given test bound state mass $M$.
The Bound state masses were located by repetitive linear 
interpolation or extrapolation to search for the point where the
eigenvalue $\Lambda$ of Eq.~(\ref{eq:EV}) is 1.  
This was repeated for each of the fermion masses
ranging from 0 to 5.0.  This procedure was used for each of the 
four non-degenerate bound state symmetries described in the appendix.

Table 2 shows the bound state masses for each of the four symmetries
(scalar ${\cal C}=+1$, scalar ${\cal C}=-1$, axi-scalar 
${\cal C}=+1$ and 
axi-scalar ${\cal C}=-1$) for all fermion masses considered.  Fig.~7a
displays the solutions $M$ for fermion mass 0--0.1.  Fig.~7b
shows $M-2m$ over the greater range of 0--5.  
The axi-scalar ${\cal C}=+1$ solution is a degenerate 
axi-scalar/axi-pseudoscalar pair of Goldstone bosons for the case
$m=0$, as seen in previous work~\cite{Bu92}.  Minor differences
between Ref.~\cite{Bu92} and the current work at $m=0$
are due to small differences in the propagator fits, as
explained in section 4.

For small $m$ the bound state masses rise rapidly with 
with increasing fermion mass.  The mass of the ``Goldstone'' 
axi-scalar ${\cal C} = +1$ state scales roughly with the square 
root of the fermion mass, in agreement with the Gell-Mann--Okubo mass formula~\cite{FS82}.  In fact, for fermion masses 0 to 0.1 
a linear regression against $\sqrt{m}$ has correlation coefficient
0.9964 with the mass growing as approximately 
$1.27 \times \sqrt{m}$.
(The accuracy of the solution at $m = 0.001$, which comes out with 
an anomalously low bound state mass, is severely affected by 
numerical inaccuracy arising from the sensitivity the bound state 
mass to the eigenvalue $\Lambda$ in Eq.~(\ref{eq:EV}).)  For large 
fermion masses, the bound state mass rises predominantly as twice 
the fermion mass plus possible logarithmic corrections.  However, 
there appears to be a good deal of noise in the large $m$ solutions, 
reflecting the difficulty in accurately modelling the fermion 
propagator deep into the timelike region from spacelike fits.  
No solutions corresponding to states of negative charge parity were 
found for $m>1.0$.  

Numerical solutions to the integral equations (\ref{eq:NREQ1}), 
(\ref{eq:NREQ1A}) 
arising from our non-relativistic treatment are listed in Table~3 
and plotted in Fig.~7b.  Solutions with positive  
$\delta$ were found for fermion masses $m \geq 1.0$ in the positive 
charge parity sector.  We were unable to locate any solutions to 
Eqs.~(\ref{eq:NREQ1}),~(\ref{eq:NREQ1A}) corresponding to 
negative charge parity states 
over a broad range of $\delta$.  Also given in Table~3 and 
Fig.~7b are the two lowest lying s-wave solutions to the 
Schr\"{o}dinger equation from the numerical work of 
Tam et al.~\cite{THY95}, given by Eq.~(\ref{eq:THY1}).  

The lack of exact agreement between the non-relativistic, 1-loop 
approximations Eqs.~(\ref{eq:NREQ1}) and (\ref{eq:NREQ1A}), and 
the Schr\"{o}dinger equation result Eq.~(\ref{eq:THY1}) is to be 
expected.  As pointed out in Section~3, a complete cancellation of 
infrared divergences can only occur if the fermion self energy is 
calculated non-perturbatively to all orders.  From Table~3, we see that 
at very high fermion masses, the accuracy of the 1-loop approximation 
is significantly affected as the conjugate poles in the propagator, 
measured in momenta scaled by the fermion mass, move closer to the 
bare fermion mass pole (see Fig.~2).  At more moderate fermion masses, 
$m\approx 5$, the 1-loop approximation is more respectable.  

We see no clear agreement between the numerical 
results of Eq.~(\ref{eq:IE}), and either non-relativistic approximation 
Eqs.~(\ref{eq:NREQ1}),~(\ref{eq:NREQ1A}), or the Schr\"{o}dinger 
equation result Eq.~(\ref{eq:THY1}).  Our analysis of the 
non-relativistic limit of the BS equation exposes the importance 
of the analytic structure of the fermion propagator in the vicinity 
of the bare fermion mass pole $p^2 = -m^2$.  The uneven nature of 
the lower two curves in Fig.~7b indicates that the determination 
of the timelike fermion propagator by an analytic fit to the spacelike 
propagator is inadequate for fermion masses $m \geq 1$.  
It is clear that a more careful analysis of the 
timelike nature of the fermion propagator, possibly involving a 
fully non-perturbative treatment of the SD equation to include 
remnant chiral symmetry breaking, is necessary for determining the 
bound state spectrum for even moderately large fermion masses.  

It is important to note that the poles in the 
fermion propagator fits listed in Table~1 lie outside the BS 
integration region $\Omega$ for all 
solutions obtained.  This can be verified by observing that all 
masses in Table~2 are lower than the values $M_{max}$ listed in Table~1.
A similar situation arises for the non-relativistic limit calculations.  
Listed in Table~3 are maximum allowed $\delta$ values if the integration 
region sampled by Eqs.~(\ref{eq:NREQ1}) and (\ref{eq:NREQ1A}) is not to 
impinge on the conjugate propagator poles $q_3^{\rm pole}$ and 
$(q_3^{\rm pole})^*$ defined in Eq. (\ref{eq:Q3POLE}).  In all cases the 
numerical results lie within the permitted region.  This requirement is 
equivalent to demanding that $q_3^{\rm pole}$ should not cross the real 
$q_3$ axis as as $\left|{\bf q}\right|$ ranges from $0$ to $\infty$.  
Interestingly, such a crossing would entail a more careful evaluation 
of residues than that carried out in Section~3 leading to the 
Schr\"{o}dinger equation.  

We note that the Schr\"{o}dinger equation results of 
Ref.~\cite{THY95} include the first five s-wave states.  
It would certainly be of interest to locate the excited states within the 
framework of our BS treatment of QED3.  
We have searched for solutions to the eigenvalue equation 
(\ref{eq:EV}) 
corresponding to excited states, and find in general no solutions 
within the mass ranges allowed by the values $M_{max}$ in Table 1.  
Since there is no reason to assume that the s-wave spectrum should 
be bounded above, it seems likely that there will be solutions to 
the BS equation for which the region of integration $\Omega$ does 
include the conjugate propagator poles discussed in Section~4.  It 
follows that the functions $f$, $U$, $V$ and $W$ in the BS 
amplitudes of these states 
should have compensating zeros, in order that the right hand side of 
the BS equation be integrable.  We conjecture that, if the fermion 
propagator has an infinite set of poles, there will be a sequence of 
excited states, the $n$th excited state having $n$ pairs of zeros in 
its BS amplitude.  This conjecture is consistent with the the first 
excited state of the Schr\"{o}dinger equation, also listed in Table~3, 
for which the wave function has a single zero.  

Although we are unable to determine accurately the spectrum in the 
large fermion mass limit, our calculations strongly suggest that 
there are no scalar or axi-scalar states with negative charge parity 
in this limit.  
This is consistent with the non-relativistic quark model in four 
dimensions in which negative charge parity scalar and pseudoscalar states are forbidden 
by the generalised Pauli exclusion principle~\cite{FS82}.  We note, 
however, that there is nothing to exclude such states in a fully 
relativistic BS treatment~\cite{LS69}, and indeed, negative charge 
parity scalar and axi-scalar states are found within the current 
model for light fermions.  

\setcounter{equation}{0}
\subsection{Conclusions}

In this paper we have solved the combination of rainbow 
Schwinger-Dyson and homogeneous Bethe-Salpeter
equations in the quenched ladder approximation for three
dimensional QED with massive fermions.  QED3
was chosen because, like QCD, it is confining but
without the complications of being non-abelian.
A four-component version of this theory is used because, also like 
QCD, it provides a parity invariant action with a spontaneously 
broken chiral-like symmetry in the massless limit.  The 
approximation is amenable to numerical solution, and should help 
assess the limitations 
of a technique frequently employed in models of QCD~\cite{BSpapers}.  

The work in this paper carries on from a previous study of the same 
subject~\cite{Bu92}, but with the following extensions.  Firstly, 
non-zero fermion masses is considered.  Secondly, an analysis of 
the fermion propagator in the complex plane is carried out in order 
to assess the appropriateness of the approximations involved.  
Thirdly, an analysis of the non-relativistic limit, i.e., large 
bare fermion mass, is made in an attempt to compare with existing 
Schr\"{o}dinger equation studies of QED3.  

The rainbow SD equation was solved in Euclidean space to give 
a fermion propagator for spacelike momenta, Euclidean $p^2 > 0$.  
The propagator is chirally asymmetric, and in the massless 
fermion limit, gives rise to a doublet of massless Goldstone positronium 
states analogous to the pion.  
Solution of the BS equation for massive positronium states 
requires knowledge of the fermion propagator $S(p)$ in the 
complex $p^2$-plane extending away from the spacelike axis, 
and a finite distance into the timelike axis $p^2 < 0$.  By 
rotating the contour of integration we were able to extend the 
spacelike solution into part of the complex plane.  However, 
the occurrence of complex conjugate poles in
the fermion propagator prevented a numerical solution to the
SD equation throughout the complete region of the complex plane sampled
by the BS equation.  This forced us to apply analytic fits to the
propagator along the positive real $p^2$-axis for use over
the required part of the complex plane.  

Our propagator fits were found to have conjugate 
poles located close to those of the direct solution for small to 
moderate fermion masses.  This,
combined with the accuracy of the fits throughout much the complex
$p^2$ plane, made our choice of propagator very attractive.
The singularities in the fits were found to move 
onto the negative real
$p^2$-axis as the fermion mass increased.  This was not
interpreted as a loss of confinement but instead attributed to
a lack of accuracy in the fits deep into the timelike region as 
the fermion mass became large.  
This reduction in accuracy of the fits for large $m$ was
due to the nature of the functions along the
positive real $p^2$-axis where the fits were made, and the presence
of a singularity near the negative real $p^2$-axis in the vicinity 
of the bare fermion mass pole $p^2 = -m^2$, but off the timelike 
axis.  

BS solutions were found for four pairs of parity degenerate states.
These pairs were the scalar/pseudoscalar ${\cal C}=+1$ and 
${\cal C}=-1$ and the 
axi-scalar/axi-pseudoscalar ${\cal C}=+1$ and ${\cal C}=-1$ states.  
For small to
moderate fermion mass the bound state mass was found to increase
smoothly with $m$.  The axi-scalar {\cal C}=+1 doublet, analogous to 
the pion, was the lowest in energy, with a mass rising roughly with 
the square root of the bare fermion mass.  For moderately large bare 
fermion masses ($m/e^2$ greater than unity) the positronium masses 
rise as twice the bare fermion mass, plus a possible logarithmic 
correction.  However, an unacceptable level of noise was found to 
develop in our results for these larger masses, which we attribute 
to inaccuracies in the analytically 
continued fermion propagators in the important region near the bare 
fermion mass pole.  No negative charge parity 
(${\cal C}=-1$) solutions were found for bare fermion masses above 
$m/e^2 \approx 1.0$, consistent with the generalised Pauli exclusion 
principle of non-relativistic QCD4.  

The conjugate poles in the fermion propagators were found 
to keep clear of the integration regions required for the BS 
solutions 
for the lowest state in each of the four space parity/charge parity 
sectors considered.  However, it appeared that this would not be so 
for any excited states.  We therefore conjecture that the excited 
positronium states have zeros in their BS amplitudes positioned so 
as to cancel the poles in the propagators encountered within the 
integral in the BS equation (\ref{eq:BS}).  This requirement
of compensating zeros was too demanding on our current numerical code,
and as a result, no excited states were found.  

In vector calculations under way at present, 
where the bound state masses are expected to be 
larger, the conjugate poles in the fermion propagator
seen in this work may interfere.  Since the fits used
in this work appear to have their
singularities close to those in the actual Schwinger-Dyson solution, 
we may find that the rainbow approximation and the resulting
propagator fits will be inadequate for a
study of vector states in QED3.  This is a very challenging problem
and we hope to report on our results in the near future.

A non-relativistic analysis of the BS equation was also carried out 
assuming, in the first instance, a 1-loop approximation to the 
fermion propagator.  However it was shown that, in order to cancel 
infrared divergences completely between the photon propagator and 
fermion self energy, as proposed by Sen~\cite{S90} and Cornwall~\cite{C80}, 
it is necessary to evaluate the fermion self 
energy non-perturbatively.  Only if this is done can the Schr\"{o}dinger 
equation be rigorously obtained in the large fermion mass limit.  
In spite of this, numerical solutions of the 1-loop equations 
give reasonable agreement with the Schr\"{o}dinger equation for 
moderately large fermion masses $m/e^2 \approx 5$.  

In summary, we were able to carry out an acceptable analysis of the bound 
state spectrum of QED3 near the chiral limit $m \rightarrow 0$ by using 
analytic fits to the spacelike fermion propagators in the BS 
Bethe-Salpeter equation, and in the non-relativistic limit 
$m \rightarrow \infty$ by expanding to lowest order in inverse powers 
of the fermion mass to obtain a Schr\"{o}dinger equation.  However, 
there remains an intermediate mass range $m/e^2 \approx 1$ for which 
neither of these techniques is adequate.  It is clear that a more careful 
non-perturbative analysis of the fermion propagator in the vicinity 
of the bare fermion mass pole is necessary before an accurate determination 
of the QED3 positronium spectrum at intermediate fermion masses can be 
made.  If a direct analogy with QCD models based on the 
Bethe-Salpeter equations is made, we conclude that particular care must 
be taken in modelling quark propagators for quarks whose mass is close to 
the mass scale of the theory, namely charm quarks.  

\renewcommand{\theequation}{\Alph{subsection}.\arabic{equation}}
\setcounter{subsection}{1}
\setcounter{equation}{0}
\subsection*{Appendix - Transformation Properties in QED3}
The four-component QED3 action in Minkowski space~\cite{Pi84}
\begin{equation}
S[A,\overline{\psi},\psi ] =
       \int \mbox{$ \, d^3x \,$} [ -\frac{1}{4} F_{\mu \nu} F^{\mu \nu} 
       +\overline{\psi} \gamma_{\mu} (i\partial^{\mu} +eA^{\mu})\psi 
       + m \overline{\psi} \psi ],
                               \label{eq:ACT}
\end{equation}
involves $4 \times 4$ matrices $\gamma_{\mu}$ which satisfy 
$\{\gamma_{\mu},\gamma_{\nu}\}=2\eta_{\mu \nu}$ where 
$\eta_{\mu \nu}={\rm diag}(1,-1,-1)$ with $\mu$ = 0, 1 and 2. 
These three matrices belong to a complete set of 16 matrices
$\{\gamma_A\}=\{I,\gamma_{4},\gamma_{5},\gamma_{45},\gamma_{\mu},
         \gamma_{\mu 4},\gamma_{\mu 5},\gamma_{\mu 45} \}$
satisfying $\frac{1}{4} {\rm tr}(\gamma_A \gamma^B) = \delta^B_A$;
$$
      \gamma_0= \left( \begin{array}{cc} \sigma_3 & 0 \\
                                0 & -\sigma_3    \end{array}\right),\;\;\;
      \gamma_{1,2}= -i\left( \begin{array}{cc} \sigma_{1,2} & 0 \\
                                0 & -\sigma_{1,2}    \end{array}\right), 
$$
$$
      \gamma_4=\gamma^4= \left( \begin{array}{cc} 0 & I \\
                                      I & 0    \end{array}\right),\;\;\;
      \gamma_5=\gamma^5= \left( \begin{array}{cc} 0 & -iI \\
                                      iI & 0    \end{array}\right),\;\;\;
      \gamma_{45}=\gamma^{45}= -i \gamma_4 \gamma_5,  
$$
$$
      \gamma_{\mu 4}=i \gamma_{\mu} \gamma_4,\;\;\;
      \gamma_{\mu 5}=i \gamma_{\mu} \gamma_5,\;\;\;
      \gamma_{\mu 45}=-i \gamma_{\mu} \gamma_4 \gamma_5,\;\;\;
      \gamma^{\mu 4,\mu 5\,{\rm or} \,\mu 45} =
      \eta^{\mu \nu}  \gamma_{\nu 4,\nu 5\,{\rm or} \,\nu 45}
$$

The three $\gamma_{\mu}$, and $\gamma_4$ and $\gamma_5$ are five mutually 
anti-commuting matrices. This is unlike the 4-dimensional case
where no analogue of $\gamma_4$ exists.  

The action Eq.~(\ref{eq:ACT}) in the massless case $m=0$
exhibits global $U(2)$ symmetry with generators 
$\{I,\gamma_{4},\gamma_{5},\gamma_{45}\}$ which is broken 
by the generation of a dynamical fermion mass \cite{DK89,Pi84} 
to a $U(1)\times U(1)$ symmetry $\{I,\gamma_{45}\}$. 
The action is also invariant with respect to discrete parity and 
charge conjugation symmetries, which for the fermion fields are
given by
\begin{equation}
\mbox{$\psi(x)$} \rightarrow \psi^\prime(x^\prime) = 
\Pi \mbox{$\psi(x)$}, \;\;\;
\mbox{$\overline{\psi}(x)$} \rightarrow \overline{\psi}^\prime
(x^\prime) = \mbox{$\overline{\psi}(x)$} \Pi^{-1},
                  \label{eq:PAR}
\end{equation}
\begin{equation}
\mbox{$\psi(x)$} \rightarrow \psi^\prime(x) = 
C \overline{\psi}(x)^{\rm T}, \;\;\;
\mbox{$\overline{\psi}(x)$} \rightarrow \overline{\psi}^\prime(x) 
= -\mbox{$\psi(x)$}^{\rm T} C^{\dagger},
                    \label{eq:CH}
\end{equation}
where $x^{\prime}=(x^0,-x^1,x^2)$. The matrices $\Pi$ and $C$ are
each determined only up to an arbitrary phase by the condition that
the action Eq.~(\ref{eq:ACT}) be invariant~\cite{Bu92}:
\begin{equation}
\Pi=\gamma_{14} e^{i\phi_P \gamma_{45}}, \;\;\;
 C=\gamma_{2} e^{i\phi_C \gamma_{45}}, (0\leq \phi_P,\phi_C < 2\pi)
\end{equation}

Scalars, pseudoscalars, axi-scalars and axi-pseudoscalars are
defined by the following transformation properties under
parity transformations

\begin{eqnarray}
\Phi^{S}(x) & \rightarrow & 
          \Phi^{S\prime}(x^{\prime}) = \Phi^{S}(x),\nonumber \\
\Phi^{PS}(x) & \rightarrow & \Phi^{PS\prime}(x^{\prime})
                             = -\Phi^{PS}(x), \nonumber \\
\Phi^{AS}(x) & \rightarrow & 
          \Phi^{AS\prime}(x^{\prime}) = R_P\Phi^{AS}(x),\nonumber \\
\Phi^{APS}(x) & \rightarrow & \Phi^{APS\prime}(x^{\prime})
                             = -R_P\Phi^{APS}(x), \label{eq:PTY} 
\end{eqnarray}
where $\Phi^{AS}$ and $\Phi^{APS}$ are doublet states 
$\Phi=(\Phi_4,\Phi_5)^T$, and
\begin{eqnarray}
R_P & = & \left(\begin{array}{cc} -\cos 2\phi_P & -\sin 2\phi_P \\
        -\sin 2\phi_P & \cos 2\phi_P  \end{array}  \right).
\label{eq:RL}
\end{eqnarray}
Similar transformation properties exist for charge conjugation.

The most general forms of the Bethe-Salpeter amplitudes~\cite{LS69} for
bound scalar and pseudoscalar states are
\begin{eqnarray}
\Gamma^S(q,P) & = & If+\not \! qg +\not \! Ph +\epsilon_{\mu \nu \rho}
                        P^{\mu}q^{\nu}\gamma^{\rho 45} k , \label{eq:GS}\\
\Gamma^{PS}(q,P) & = & \gamma_{45} \Gamma^S(q,P), \label{eq:GP}
\end{eqnarray}
where $f,g,h$ and $k$ are functions only of $q^2,P^2$ and 
$q\cdot P$.   
BS amplitudes corresponding to the components $\Phi_4$ and 
$\Phi_5$ of 
axi-scalars and axi-pseudoscalars take the general form
\begin{equation}
  \left(\begin{array}{c} \Gamma^{(4)}(q,P) \\ \Gamma^{(5)}(q,P)
        \end{array} \right)^{AS} = 
\left(\begin{array}{c} \gamma_4 \\
           \gamma_5\end{array} \right)f + 
          \left(\begin{array}{c} \gamma_{\mu 4} \\ 
     \gamma_{\mu 5}\end{array} \right) (q^{\mu}g+P^{\mu}h)
+\epsilon_{\mu \nu \rho} P^{\mu} q^{\nu}
  \left(\begin{array}{c} \gamma^{\rho 5} \\
              - \gamma^{\rho 4}\end{array} \right)k,  \label{eq:GAS}
\end{equation}
and 
\begin{equation}
  \left(\begin{array}{c} \Gamma^{(4)}(q,P) \\ \Gamma^{(5)}(q,P)
        \end{array} \right)^{APS} = \gamma_{45} 
\left(\begin{array}{c} \Gamma^{(4)}(q,P) \\ \Gamma^{(5)}(q,P)
        \end{array} \right)^{AS}.
 \label{eq:GAPS}
\end{equation}

Furthermore, the charge parity ${\cal C}=\pm 1$ of the bound states 
is determined by
the parity of the functions $f,g,h$ and $k$ under the transformation
$q\cdot P \rightarrow -q\cdot P$.  The quantity 
$q\cdot P$ is the only Lorentz invariant which 
changes sign under charge conjugation and thus determines the
charge parity of those functions.

Our conventions for Euclidean space quantities are summarised in
Appendix A of Ref.~\cite{Bu92}.  In particular Euclidean momenta and 
Dirac matrices are defined by
$$
P_3^{({\rm E})} = -iP_0^{({\rm M})},\hspace{5 mm}
P_{1,2}^{({\rm E})} = P_{1,2}^{({\rm M})},\hspace{5 mm}
\gamma_3^{({\rm E})} = \gamma_0^{({\rm M})}, \hspace{5 mm}
\gamma_{1,2}^{({\rm E})} = i\gamma_{1,2}^{({\rm M})}.
$$
\subsection*{Acknowledgments}
We are grateful to C. J. Hamer, A. Tam and P. Maris for helpful 
discussions, and the National Centre for Theoretical Physics at 
the Australian National University for hosting the Workshop on 
Non-Perturbative Methods in Field Theory where part of this work was 
completed.  
\pagebreak

\pagebreak
\begin{figure}[ht]
\begin{center}
\begin{tabular}{|c|c|c|}\hline
\rule{0mm}{5mm} {$m$} & {$p^2$} & {$M_{max}$} \\ \hline
\rule{0mm}{5mm} 0.000 & $-$0.0034 $\pm i $ 0.0057 & 0.142 \\\hline
\rule{0mm}{5mm} 0.001 & $-$0.0041 $\pm i $ 0.0064 & 0.153 \\ \hline
\rule{0mm}{5mm} 0.004 & $-$0.0060 $\pm i $ 0.0086 & 0.182 \\ \hline
\rule{0mm}{5mm} 0.009 & $-$0.0081 $\pm i $ 0.0140 & 0.206 \\ \hline
\rule{0mm}{5mm} 0.016 & $-$0.0121 $\pm i $ 0.0192 & 0.247 \\ \hline
\rule{0mm}{5mm} 0.025 & $-$0.0216 $\pm i $ 0.0260 & 0.325 \\ \hline
\rule{0mm}{5mm} 0.036 & $-$0.0314 $\pm i $ 0.0345 & 0.386 \\ \hline
\rule{0mm}{5mm} 0.049 & $-$0.0468 $\pm i $ 0.0387 & 0.464 \\ \hline
\rule{0mm}{5mm} 0.064 & $-$0.0618 $\pm i $ 0.0417 & 0.522 \\ \hline
\rule{0mm}{5mm} 0.081 & $-$0.0815 $\pm i $ 0.0440 & 0.590 \\ \hline
\rule{0mm}{5mm} 0.1 & $-$0.0647 $\pm i $ 0.0000 & 0.509 \\ \hline
\rule{0mm}{5mm} 0.5 & $-$0.4894 $\pm i $ 0.0000 & 1.399 \\ \hline
\rule{0mm}{5mm} 1 & $-$1.4260 $\pm i $ 0.0000 & 2.388 \\ \hline
\rule{0mm}{5mm} 2 & $-$4.8925 $\pm i $ 0.0000 & 4.424 \\ \hline
\rule{0mm}{5mm} 3 & $-$10.3776 $\pm i $ 0.0000 & 6.443 \\ \hline
\rule{0mm}{5mm} 4 & $-$17.9613 $\pm i $ 0.0000 & 8.476 \\ \hline
\rule{0mm}{5mm} 5 & $-$27.3616 $\pm i $ 0.0000 & 10.462 \\ \hline
\end{tabular}
\end{center}
\end{figure}
\begin{center}
\parbox{140mm}{Table 1: Conjugate singularities for fermion propagator fit and corresponding limits on bound state mass
for fermion masses from 0 to 5.0.}
\end{center}
\pagebreak
\begin{figure}[ht]
\begin{center}
\begin{tabular}{|c|c|c|c|c|}\hline
\rule{0mm}{5mm}
       {$m$} & {Scalar ${\cal C}=+1$} & {Scalar ${\cal C}=-1$} & 
       {Axi-scalar ${\cal C}=+1$} & {Axi-scalar ${\cal C}=-1$} \\ 
       \hline
\rule{0mm}{5mm} 0 [Ref.~\cite{Bu92}] & 0.080 $\pm$ 0.001 & 0.123 
$\pm$ 0.002 & 0 & 0.111 $\pm$ 0.002  \\ \hline
\rule{0mm}{5mm}   0   & 0.077 & 0.118 & 0 & 0.108 \\ \hline
\rule{0mm}{5mm} 0.001 & 0.087 & 0.126 & 0.004 & 0.116 \\ \hline
\rule{0mm}{5mm} 0.004 & 0.110 & 0.151 & 0.054 & 0.140 \\ \hline
\rule{0mm}{5mm} 0.009 & 0.140 & 0.178 & 0.090 & 0.167 \\ \hline
\rule{0mm}{5mm} 0.016 & 0.175 & 0.217 & 0.127 & 0.204 \\ \hline
\rule{0mm}{5mm} 0.025 & 0.215 & 0.269 & 0.167 & 0.254 \\ \hline
\rule{0mm}{5mm} 0.036 & 0.256 & 0.316 & 0.208 & 0.300 \\ \hline
\rule{0mm}{5mm} 0.049 & 0.298 & 0.367 & 0.248 & 0.350 \\ \hline
\rule{0mm}{5mm} 0.064 & 0.343 & 0.411 & 0.293 & 0.390 \\ \hline
\rule{0mm}{5mm} 0.081 & 0.389 & 0.456 & 0.340 & 0.431 \\ \hline
\rule{0mm}{5mm}  0.1  & 0.439 & 0.496 & 0.391 & 0.479 \\ \hline
\rule{0mm}{5mm}  0.5  & 1.311 & 1.388 & 1.261 & 1.352 \\ \hline
\rule{0mm}{5mm}   1   & 2.297 & 2.387 & 2.243 & 2.336 \\ \hline
\rule{0mm}{5mm}   2   & 4.330 &   -   & 4.233 &   -   \\ \hline
\rule{0mm}{5mm}   3   & 6.348 &   -   & 6.227 &   -   \\ \hline
\rule{0mm}{5mm}   4   & 8.379 &   -   & 8.243 &   -   \\ \hline
\rule{0mm}{5mm}   5   & 10.365 &  -   & 10.219 &   -   \\ \hline
\end{tabular}
\end{center}
\end{figure}
\begin{center}
\parbox{140mm}{Table 2: Bound state masses for fermion masses from 0 to 5.0.  
(All masses $\pm 0.001$ unless otherwise stated).  The Axi-scalar ${\cal C}=+1$
solution with $m=0$ stated here is an analytic result.}
\end{center}
\pagebreak
\begin{figure}[ht]
\begin{center}
\begin{tabular}{|c|c|c|c|c|c|}\hline
\rule{0mm}{5mm} {$m$} & {$\delta_{max}$} & {Scalar} & 
{Axi-scalar} & {Eq.(\ref{eq:THY1})} with  & {Eq.(\ref{eq:THY1})} 
with \\ 
\rule{0mm}{5mm}  &  &  {Eq.(\ref{eq:NREQ1})} & 
{Eq.(\ref{eq:NREQ1A})} & {$\lambda=\lambda_{0}$} & 
{$\lambda=\lambda_{1}$}  \\ \hline
\rule{0mm}{5mm} 1 & 0.332 & 0.285 & 0.262 & 0.322 & 0.503 \\ \hline
\rule{0mm}{5mm} 2 & 0.421 & 0.371 & 0.338 & 0.377 & 0.558 \\ \hline
\rule{0mm}{5mm} 3 & 0.473 & 0.419 & 0.381 & 0.409 & 0.590 \\ \hline
\rule{0mm}{5mm} 4 & 0.511 & 0.452 & 0.410 & 0.432 & 0.613 \\ \hline
\rule{0mm}{5mm} 5 & 0.540 & 0.476 & 0.433 & 0.450 & 0.631 \\ \hline
\rule{0mm}{5mm} 100 & 0.947 & 0.792 & 0.734 & 0.688 & 0.869 \\ 
\hline
\rule{0mm}{5mm} 1000 & 1.272 & 1.030 & 0.968 & 0.872 & 1.052 \\ 
\hline
\end{tabular}
\end{center}
\end{figure}
\begin{center}
\parbox{140mm}{Table 3: Non-relativistic $\delta$ solutions for
positive charge parity, from Eqs.~(\ref{eq:NREQ1}), 
(\ref{eq:NREQ1A}) and the two lightest s-wave 
Schr\"{o}dinger equation results of Tam et al.~\cite{THY95}, 
Eq.~(\ref{eq:THY1}) using $\lambda_{0}=1.7969$ and
$\lambda_{1}=2.9316$.  The first column contains the maximum
values of $\delta$ allowed before conjugate singularities
arise in the fermion propagator used in our non-relativistic
calculations.}
\end{center}

\pagebreak
\section*{Figures}
\begin{description}
  \vspace*{5mm}
  \item[Figure 1:] Diagrammatic representation of Eq.~(\ref{eq:BS}).
  \vspace*{5mm}
  \item[Figure 2:] Figures 2a and 2b show 1-loop approximations 
using eqs.(\ref{eq:SIGDEF}), (\ref{eq:ABDEF}) and 
(\ref{eq:A1LOOP}) -- (\ref{eq:SIGB1}) 
for $m^2\sigma_V$ and $m\sigma_S$ respectively (solid lines).  
These are compared with the vector and scalar parts of the approximation 
Eq.~(\ref{eq:SAPX}) (dashed lines).  
The curves are drawn for fermion masses (from bottom to top) 
$1$, $2$, $4$, $8$, and $\infty$.
  \vspace*{5mm}
  \item[Figure 3:] The first and second (deformed) contours of 
integration $C_1$ and $C_2$ for solution to Eq.~(\ref{eq:ABINTS}). 
  \vspace*{5mm}
  \item[Figure 4:] Figure 4a compares the SDE solutions
and fitting functions for $A-1$ for fermion masses (from top to bottom) 
$m=$0, 0.025, 0.1, 1 and 5.  Figure 4b shows $B-m$
for fermion masses (from bottom to top) $m=$0, 0.025, 0.1, 1 and 5.
  \vspace*{5mm}
  \item[Figure 5:] Figures 5a and 5b show SDE solutions and function
fits for $\sigma_V$ and $\sigma_S$ respectively for fermion 
mass $0$ with angles $\phi=0$ ($\Diamond$), $\frac{\pi}{8}$ ($+$) 
and $\frac{\pi}{4}$ ($\Box$).
  \vspace*{5mm}
  \item[Figure 6:] Figure 6a compares the function $A(p^2)$ from 
the rainbow SDE calculation (solid curve) and the 1-loop result 
(dashed curve) and figure 6b compares $B(p^2)$ results along the positive real $p$ axis for fermion mass $m=5.0$.
  \vspace*{5mm}
  \item[Figure 7:] Figure 7a shows bound state masses ($M$) against
fermion mass 0--0.1.  Figure 7b is a plot of $M-2m$ for $m=$0--5. 
In each plot the scalar ${\cal C}=+1$ ($\Diamond$), 
scalar ${\cal C}=-1$ ($+$), axi-scalar ${\cal C}=+1$ 
($\Box$) and axi-scalar ${\cal C}=-1$ ($\times$) states are drawn 
with solid curves.  The non-relativistic predictions of Eq.~(\ref{eq:NREQ1}) 
and Eq.~(\ref{eq:NREQ1A}) are the scalar ${\cal C}=+1$ ($\Diamond$)
and axi-scalar ${\cal C}=+1$ ($\Box$) states respectively and are
drawn with dashed lines.
Eq.~(\ref{eq:THY1}) with $\lambda=\lambda_0$ (lower solid curve with no
symbols) and with $\lambda_1$ (upper solid curve with no symbols)
are also plotted in figure 7b.
  \vspace*{5mm}
\end{description}

\begin{thebibliography}{99}
%
\bibitem{BPR92} C.\ J.\ Burden, J.\ Praschifka.\ and.\  
C.\ D.\ Roberts,  Phys.\ Rev.
                        {\bf D46} (1992) 2695.
%
\bibitem{BSpapers} J. Praschifka, C. D. Roberts, R. T. Cahill, Intern. J.
Mod. Phys. A 4 (1989) 4929; Y.-b. Dai, C.-s. Huang and D.-s. Liu, Phys. Rev.
D 43 (1991) 1717; K.-I. Aoki, T. Kugo and M. G. Mitchard, Phys. Lett. B 266
(1991) 467; H. J. Munczek and P. Jain, Phys. Rev. D 46 (1992) 438;
P. Jain and H. J. Munczek, Phys. Rev. D 48 (1993) 5403;  
C.\ J.\ Burden, et al.\, {\it Separable approximation to the Bethe-Salpeter 
in QCD}, proceedings of the Lattice '95 Conference, 1995 (to appear); 
R. T. Cahill and S. T. Gunner, {\it Quark and gluon propagators from 
meson data}, Flinders University preprint, 1995.    
%
\bibitem{SC94} S.\ J.\ Stainsby, and.\  
  R.\ T.\ Cahill,  Mod.\ Phys.\ Lett. {\bf A9} (1994) 3551.  
%
\bibitem{Bu92} C.\ J.\ Burden, Nucl.\ Phys.
                        {\bf B387} (1992) 419.
%
\bibitem{Ma93} P.\ Maris, {\it Nonperturbative Analysis of the Fermion 
Propagator: Complex Singularities and Dynamical Mass Generation},  PhD 
thesis, (1993).
%
\bibitem{Ma95} P.\ Maris, {\it Confinement and complex singularities in 
QED3}, University of Nagoya pre-print, 1995.
%
\bibitem{DS79} R.\ Delbourgo, and.\  M.\ Scadron,  J.\ Phys.
                        {\bf G5} (1979) 1621.
%
\bibitem{YH91} C. M. Yung and C. J. Hamer, Phys. Rev. {\bf D44} (1991) 2595.
%
\bibitem{THY95} A.\ Tam, C.\ J.\ Hamer and C.\ M.\ Yung, {\it Light-cone 
quantisation approach to quantum electrodynamics in (2+1) dimensions}, 
                            UNSW preprint  PRINT-94-0182, 1994; see also 
               V.\ G.\ Koures, {\it Solving the Coulomb Schr\"{o}dinger 
               equation in $d = 2+1$ via sinc collocation}, Univ. of 
               Utah preprint UTAH-IDR-CP-05.
%
\bibitem{S90} D.\ Sen,  Phys.\ Rev.\ {\bf D41} (1990) 1227.
%
\bibitem{C80} J.\ M.\ Cornwall,  
Phys.\ Rev.\ {\bf D22} (1980) 1452.
%
\bibitem{RW94} C.\ D.\ Roberts and A.\ G.\ Williams,  Prog.\ 
Part.\ and.\ Nucl.\ Phys.\ {\bf 33} (1994) 475.
%
\bibitem{SC92} S.\ J.\ Stainsby, and.\  R.\ T.\ Cahill,  
      Int.\ J.\ Mod.\ Phys. {\bf A7} (1992) 7541.
%
\bibitem{FS82} D.\ Flamm and F.\ Sch\"{o}berl, {\it Introduction to the
Quark Model of Elementary Particles}, Gordon and Breach, 1982.  
%
\bibitem{LS69} C.\ H.\ Llewellyn Smith, Ann.\ Phys.\ {\bf 53} (1969) 521,
                  Llewellyn Smith's vertex $\chi$ is related to our
                  vertex $\Gamma$ via: 
                  $\chi(\frac{1}{2}P,q) = S(\frac{1}{2}P+q)
                   \Gamma(q,P) S(\frac{1}{2}P-q)$.
%
\bibitem{Pi84} R.\ D.\ Pisarski, Phys.\ Rev.\ {\bf D29} (1984) 2423.
%
\bibitem{DK89} E.\ Dagotto, J.\ B.\ Kogut and A.\ Kocic,
                 Phys.\ Rev.\ Lett.\ {\bf 62} (1989) 1083,
                 Nucl.\ Phys.\ {\bf B334} (1990) 279.
%
\end{thebibliography}
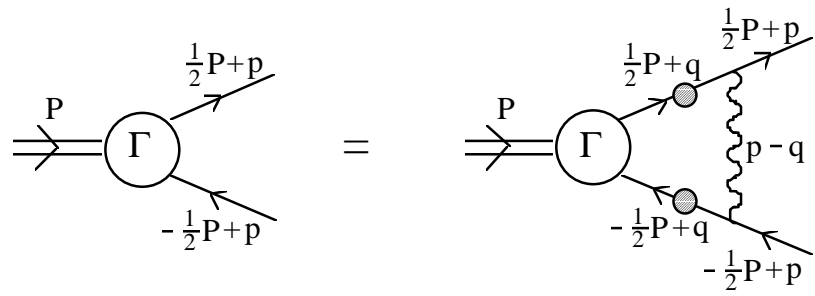
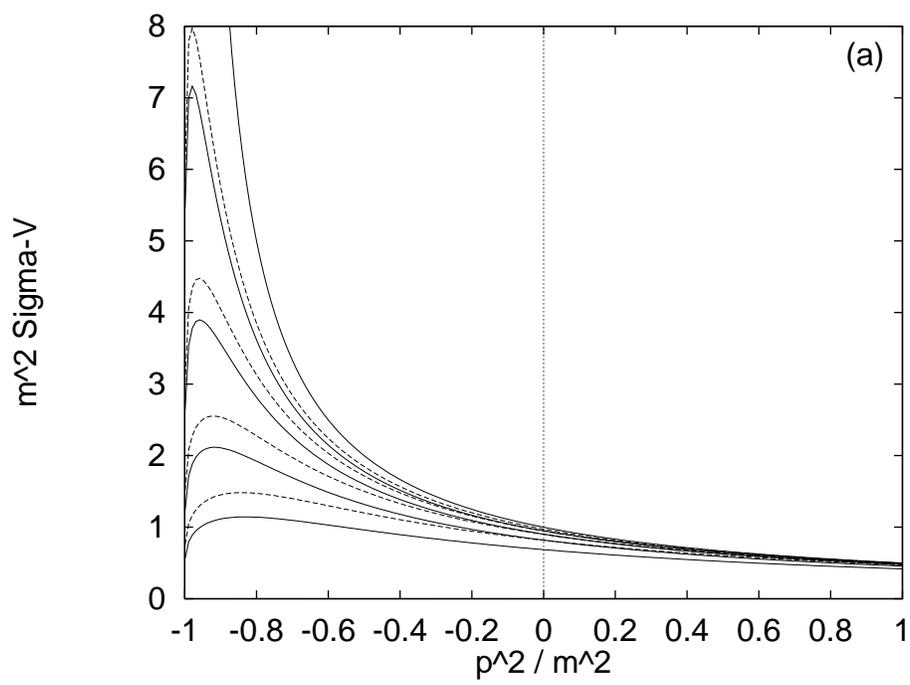
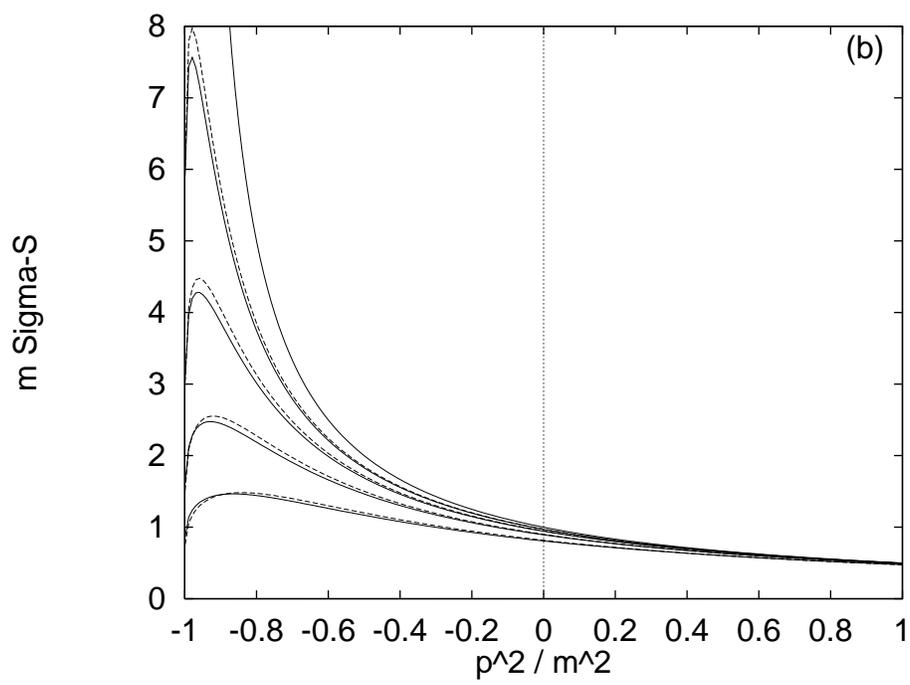
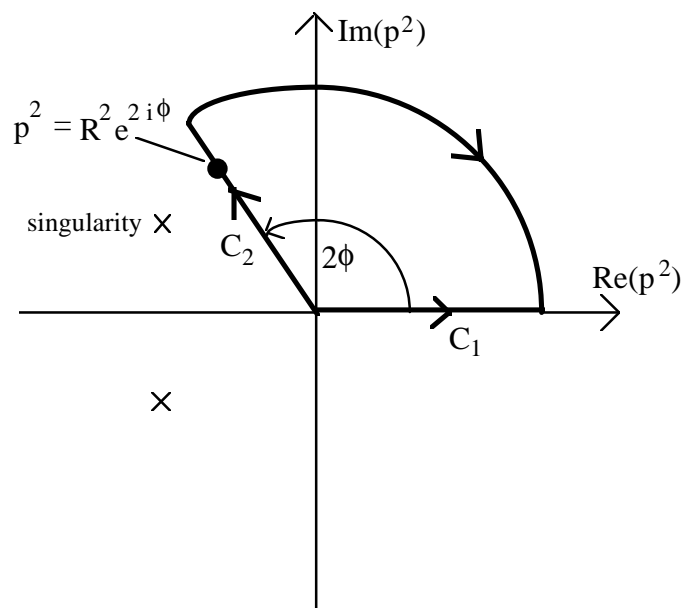
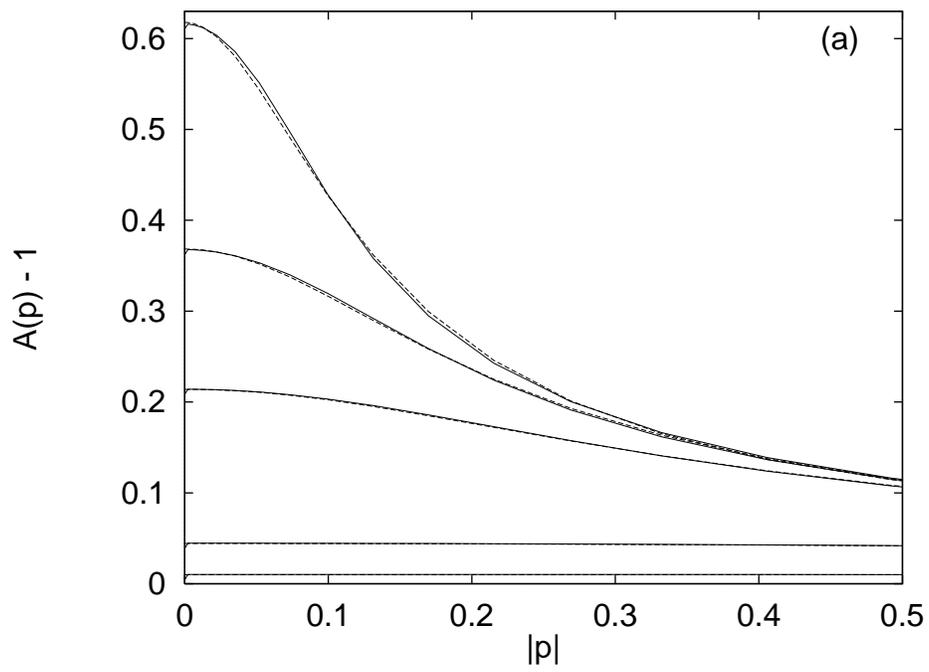
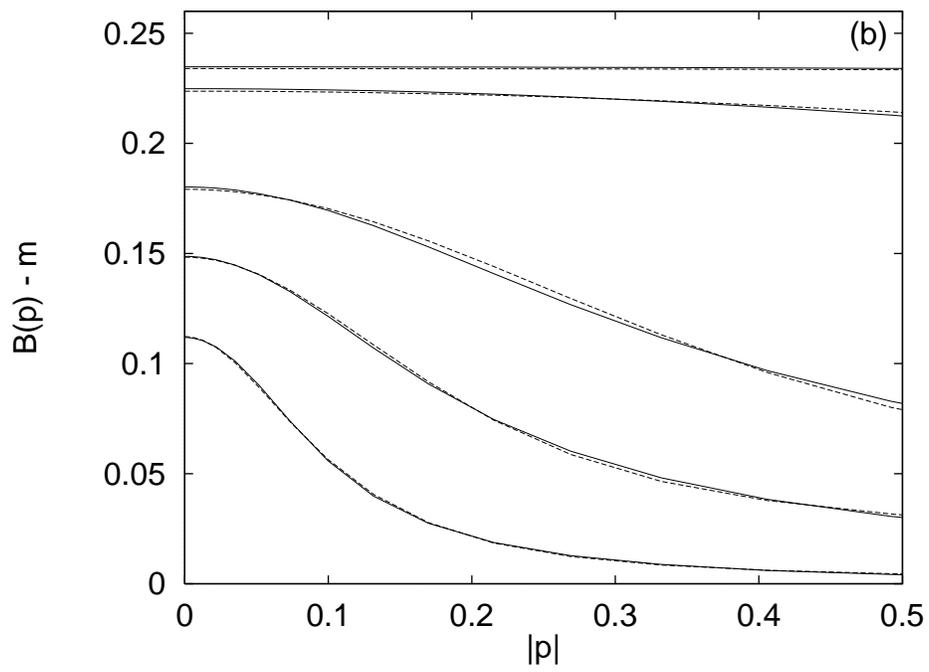
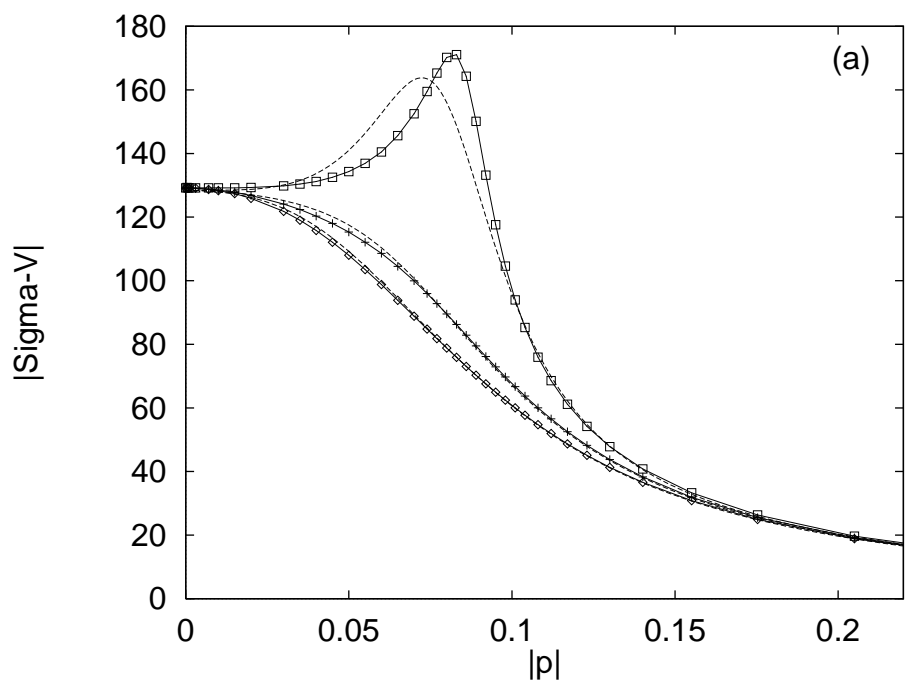
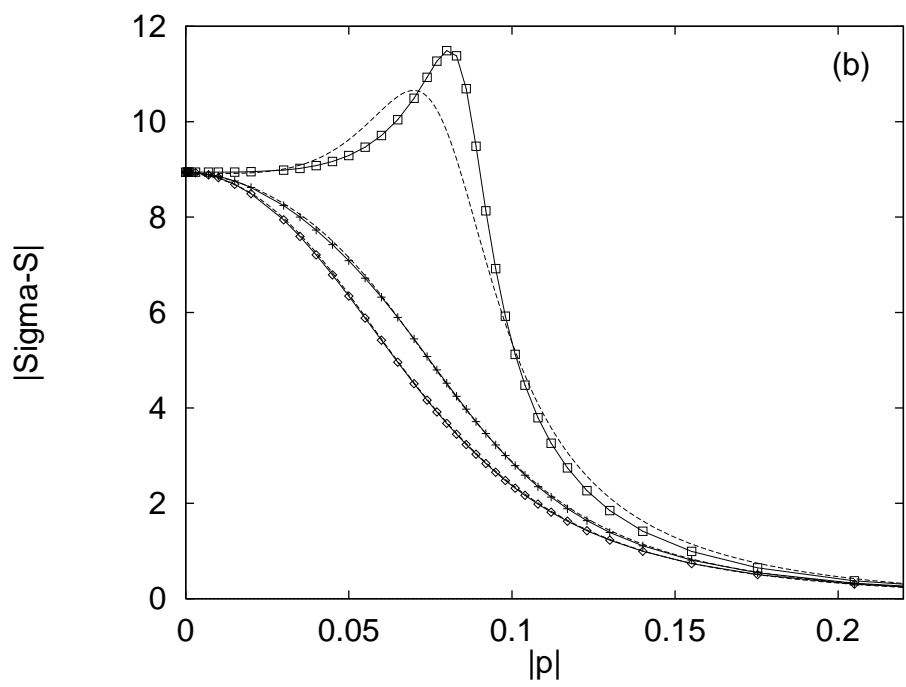
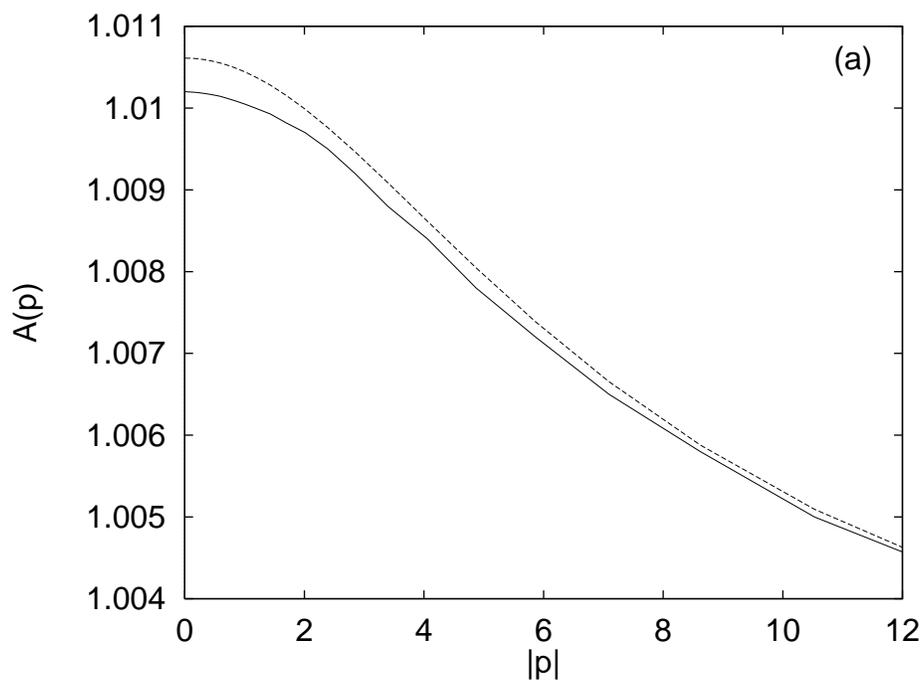
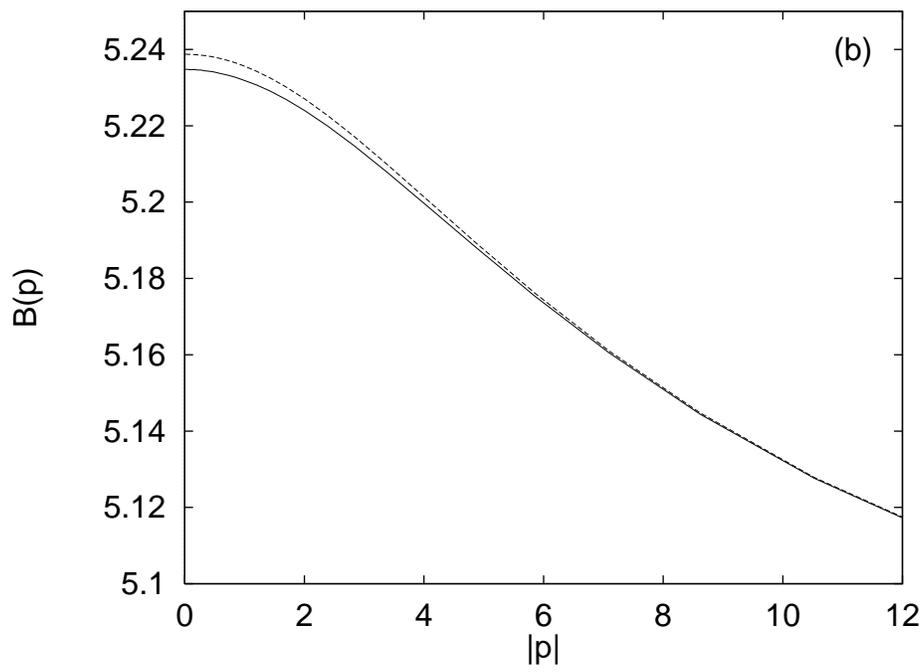
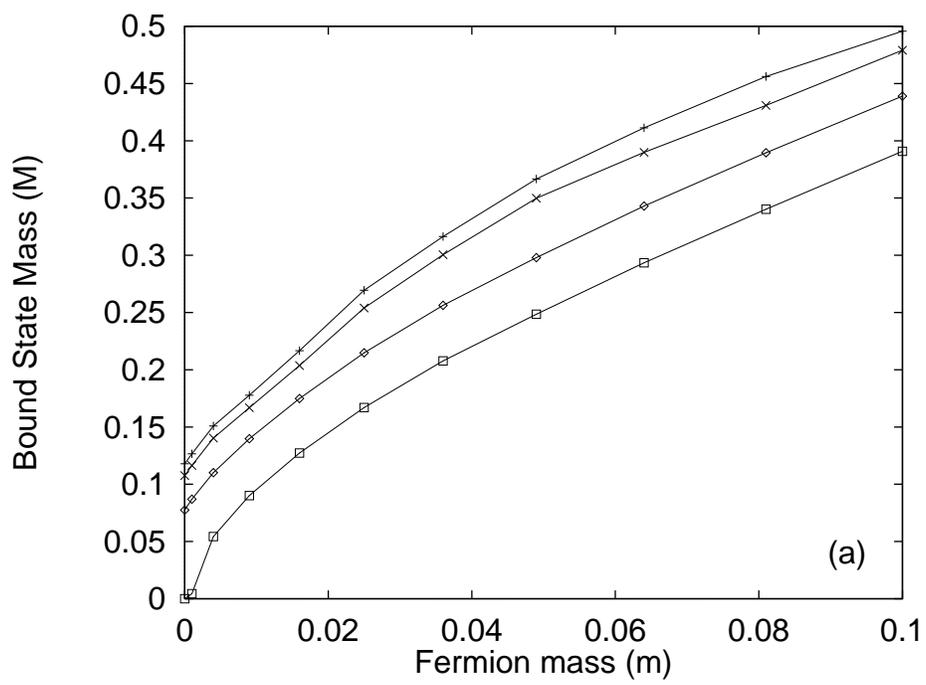
\end{document}